\documentclass[fleqn,usenatbib]{mnras}

\usepackage[T1]{fontenc}
\usepackage{soul}
\usepackage{longtable}
\usepackage{pifont}

\usepackage{graphicx}	
\usepackage{amsmath}	
\usepackage{amssymb}	
\usepackage{wasysym}

\usepackage{newtxtext,newtxmath}
\usepackage{hyperref}

\title[Detection of periodic spin period reversals in Vela X-1]{Detection of nearly periodic spin period reversals in Vela X-1 on long time-scales: inkling of solar-like cycle in the donor star?}

\author[A. D. Chandra et al.]{
Amar Deo Chandra,$^{1,2}$\thanks{E-mail: amar.deo.chandra@gmail.com}
Jayashree Roy,$^{3,1}$
P. C. Agrawal$^{1,4}$
and Manojendu Choudhury$^{5,1}$
\\
$^{1}$UM-DAE Centre for Excellence in Basic Sciences, University of Mumbai, Vidyanagari Campus, Kalina, Santacruz (East), Mumbai, Maharashtra 400098, India\\
$^{2}$Center of Excellence in Space Sciences India, Indian Institute of Science Education and Research Kolkata, Mohanpur 741246, West Bengal, India\\
$^{3}$Inter-University Center for Astronomy and Astrophysics, Post Bag 4, Pune, Maharashtra 411007, India\\
$^{4}$ Senior Professor (Retd.), Dept. of Astronomy and Astrophysics, Tata Institute of Fundamental Research, Homi Bhabha Road, Mumbai 40005, India\\
$^{5}$ Department of Physics, St. Xavier's College (Autonomous), 5 Mahapalika Marg, Mumbai 400001, Maharashtra, India
}

\date{Accepted XXX. Received YYY; in original form ZZZ}

\pubyear{2021}

\begin{document}
\label{firstpage}
\pagerange{\pageref{firstpage}--\pageref{lastpage}}
\maketitle

\begin{abstract}
We explore the long-term evolution of the spin period of the High Mass X-ray Binary (HMXB) pulsar Vela X-1 over a period of 46 yr. Our analysis indicates nearly periodic variations in the spin period of the pulsar on time-scales of about 5.9 yr. There is suggestion of an overall spin-down behaviour of the pulsar though it is noticed that the source appears to stay near its equilibrium period 283.4 s since MJD 51000, with rather erratic spin-up/spin-down episodes around this value. Our study suggests nearly cyclic turnover in the spin behaviour of the pulsar from spin-up to spin-down regimes on time-scales of about 17-19 yr. To our knowledge this is the first report of periodic variation in the spin behaviour of a wind-fed accreting pulsar. We also observe erratic episodes of spin-up and spin-down behaviour on relatively shorter time-scales which is a well known archetype
of this wind-fed X-ray pulsar. We investigate whether nearly periodic long-term spin period changes in the pulsar can be explained by using known mechanisms of torque reversals in the accretion powered neutron stars. It appears that changes in the accretion environment of the pulsar using current ideas can probably lead to long-term spin period changes in this X-ray pulsar.\\
\end{abstract}




\section{Introduction}

Vela X-1 (4U 0900-40) is an eclipsing High mass X-ray binary (HMXB) discovered during rocket borne X-ray observations in 1967 \citep{chodil1967x}. It is located at a distance of $\sim$2.0 kpc \citep{sadakane1985ultraviolet,nagase1989accretion} in the Vela constellation. Recent estimates using \textit{Gaia} data infer distance of $2.42^{+0.19}_{-0.17}$ kpc \citep{bailer2018}. The system consists of a massive B0.5Ib supergiant HD 77581 \citep{hiltner1972binary,brucato1972optical,vidal1973hd,jones1973optical} having mass of about $\sim$ 23 $\rm{M_\odot}$ and radius of $\sim$ 34 $\rm{R_\odot}$ \citep{van1976mass,joss1984neutron,nagase1989accretion,van1995spectroscopy} and a neutron star with mass $\sim$ 1.8 $\rm{M_\odot}$ \citep{van1976mass,nagase1989accretion,barziv2001mass,rawls2011ref}. The orbital period of the binary system is about nine days \citep{hiltner1972binary,forman1973uhuru,vidal1973hd,watson1977ariel,van1995spectroscopy}. Due to the close proximity of about 1.7 $\rm{R_\star}$ \citep{conti1978stellar,quaintrell2003mass} between the neutron star and its companion, the neutron star is immersed in the dense stellar wind of the donor star having typical mass loss rate of about $\dot{M} \mathrm{\sim 10^{-6} ~{M_\odot} \,yr^{-1}}$ \citep{hutchings1974x,dupree1980simultaneous,nagase1986circumstellar,sako1999x}. A fraction of the stellar wind is captured and channelled along the strong magnetic field ($\sim 2.7 \times 10^{12}$\,G; \citep{kretschmar1996absorption,kreykenbohm2002confirmation,coburn2002magnetic} of the neutron star onto the magnetic poles, producing regular X-ray pulsations caused by the spin period $\sim 283$\,s \citep{rappaport1975discovery,mcclintock1976discovery} of the neutron star. Although Vela X-1 is known to be a persistent source having luminosity of about ${4}\times 10^{36}$\,erg\,s$^{-1}$ \citep{mccray1984spectral,sadakane1985ultraviolet,nagase1986circumstellar,kreykenbohm2002confirmation}, it shows a plethora of X-ray variabilities such as sudden flares lasting a few minutes to several hours wherein the luminosity increases by several folds within very short time-scales of a few tens of seconds \citep{lapshov1992two,staubert2004integral,kreykenbohm2008high}. Occurrence of sudden flares in this system are not so well understood and is believed to be due to enhanced accretion rate due to variabilities in the stellar wind from the companion star \citep{nagase1983light,haberl1990x} or accretion of clumpy stellar wind \citep{staubert2004integral,ducci2009,furst2010x,odaka2013short}. Some studies suggest that sudden flares might be related to formation of transient accretion disc \citep{inoue1984sudden,taam1989numerical,haberl1990x,kreykenbohm2008high}. Another bizzare manifestation seen in Vela X-1 is occurrence of abrupt ``off-states'' wherein X-ray pulsations cessation (within less than the pulse period) is observed for several tens of minutes at a time \citep{inoue1984sudden,lapshov1992two,kreykenbohm1999,kreykenbohm2008high,doroshenko2011witnessing,sidoli2015probing}. These states are poorly understood and might be caused by changes in the accretion rate due to variabilities in the stellar wind \citep{lapshov1992two,coburn2002magnetic}. Some  earlier studies also suggest that ``off-states''  might be associated with formation of transient accretion discs \citep{inoue1984sudden} or the accretion is choked due to the sudden onset of propeller effect \citep{kreykenbohm2008high}. It has also been suggested that the onset of these ``off-states'' can be caused due to transition from the higher luminosity Compton cooling regime to the lower luminosity radiative cooling regime \citep{shakura2013nature} or due to unstable hydrodynamic flows in the vicinity of the neutron star \citep{manousakis2015origin}. Recent numerical studies suggest formation of temporary accretion discs in wind-fed X-ray pulsars \citep{el2019formation,el2019wind,karino2019stellar} but conclusive evidence of their existence has been elusive. Interestingly, \citet{liao2020spectral} infer presence of temporary accretion disc in Vela X-1 during an extended low state lasting at least 30 ks which was accompanied by unusual spin-up event and similar Fe K $\alpha$ fluxes compared to the preceding flaring period. \\

Long-term monitoring of the spin period of Vela X-1 since 1975 has shown erratic spin period variations over time which is an archetype of wind-fed X-ray pulsars \citep{deeter1989vela,bildsten1997observations}. Spin excursions on short time-scales are considered to be caused by the dynamic accretion torque acting on the neutron star as a result of changes in the stellar wind from the donor star  \citep{kreykenbohm2008high}. Vela X-1 is known to show variable episodes of spin-up and spin-down \citep{nagase1981spin}. The evolution of spin period is most appropriately described by a random walk model \citep{tsunemi1989all,ziolkowski1985rotational}. Although the source shows strong pulse to pulse variations which is a tell-tale manifestation of fluctuating rate of accretion \citep{staubert1980hard,nagase1984secular,kretschmar1997phase}, pulse profiles folded over several pulse periods show remarkable stability (for 10 pulses or more; \citet{staubert1980hard}, even over decades  \citep{raubenheimer1990pulsed}). This suggests that the magnetic environment of the neutron star is very stable on long time-scales. Recent measurements of surface magnetic fields of X-ray pulsars using detection of cyclotron resonance scattering effects suggest that the magnetic field of neutron stars are almost constant on very long time-scales of about $10^8$ yr  \citep{makishima1999cyc}. However, recent studies have found changes in cyclotron line energies in Vela X-1 on long time-scales using the \textit{Neil Gehrels Swift Observatory}/BAT observations \citep{la2016swift,ji2019long}. This suggests that the inferred magnetic field decays at the rate of about $3\times10 ^{10}$\,G\, yr$^{-1}$ \citep{la2016swift}. The most recent spin evolution study of Vela X-1 found the pulsar showing spin-down behaviour \citep{kreykenbohm1999}.\\

Investigations of spin period variations on various time-scales (few tens of days to several hundred days) have been carried out in several X-ray pulsars and at least one torque reversal has been detected in some of these pulsars (Cen X-3 \citep{tsunemi1989pulse}; several pulsars studied by \citet{nagase1989accretion} and 
 references therein; OAO 1657-415 \citep{chakrabarty1993discovery}; several pulsars monitored and studied by the Burst and Transient Source Experiment (BATSE) \citep{bildsten1997observations}; 4U 1626–67 \citet{wilson1993observations}, \citet{chakrabarty1997torque}, \citet{camero2009new}; GX 1+4 \citet{makishima1988spin}, \citet{chakrabarty1997correlation}, \citet{gonzalez2012spin}; 4U 1907+09 \citet{fritz2006torque}, \citet{inam2009recent}; 4U 0114+650 \citep{hu2016evolution}; LMC X-4 \citep{molkov2016near}; NGC 300 ULX1 \citep{vasilopoulos2019ngc}; 2S 1845-024 and several other pulsars monitored and studied by the Gamma-ray Burst Monitor (GBM) aboard the \textit{Fermi} Gamma-ray Space Telescope \citep{malacaria2020ups}). A long-term periodicity of about 9 yr and 6.8 yr has been detected in the spin evolution of Cen X-3 \citep{tsunemi1989pulse} and LMC X-4 \citep{molkov2016near} respectively. It is intriguing to note that GX 1+4, which is a disc-fed pulsar, has shown only one transition from secular spin-up to monotonic spin-down behaviour \citep{makishima1988spin,chakrabarty1997correlation,gonzalez2012spin} during its spin evolution monitored over almost five decades. Cen X-3 has a massive O6-8 III type companion star V779 Cen \citep{krzeminski1974identification} which has a strong wind. However, presence of an accretion disc has been detected in this system \citep{tjemkes1986optical} which suggests that accretion induced spin changes in this pulsar are mainly driven by accretion from the disc. This makes Cen X-3 a predominantly disc-fed pulsar. The companion star of LMC X-4 is an O8 III type massive star \citep{kelley1983discovery,falanga2015ephemeris} and the neutron star accretes from an accretion disc in this system \citep{lang1981discovery} making this a disc-fed source. It should be noted that though Cen X-3 and LMC X-4 are disc-fed pulsars unlike Vela X-1 which is a wind-fed pulsar, they also exhibit long-term periodicity in their spin evolution. This makes Vela X-1 to be the first wind-fed X-ray pulsar where such a long-term periodicity (on time-scales of years) in the spin evolution has been detected.\\

In this paper, we investigate the long-term ($\sim$ 46 yr) spin evolution of Vela X-1 using spin period measurements from 24 different observatories. Most of the spin period measurements are taken from literature and we deduce spin periods, not reported earlier, from using the \textit{Rossi X-ray Timing Explorer (RXTE)}/Proportional Counter Array (PCA) and the \textit{AstroSat}/Large Area X-ray Proportional Counter (LAXPC) observations. The complete list of spin period measurements is shown in the appendix \ref{appendix:a}. The paper is organized as follows. After the introduction, in section 2 we first describe  observations and data analysis from archival \textit{RXTE}/PCA observations of Vela X-1. Then we describe recent observations from the \textit{AstroSat} mission and LAXPC data analysis procedures. In section 3 we describe our main results related to the detection of long-term spin-down trend in Vela X-1 and detection of nearly periodic spin period reversals on long time-scales. In section 4 we discuss possible changes in the accretion landscape of this wind-fed X-ray pulsar using current ideas which can explain nearly periodic long-term torque reversals. A summary of all our findings is presented in section 5.   

\section{Observations and data analysis}
The main goal of our work is to explore the long-term spin period evolution of Vela X-1 since its discovery almost five decades ago. To the best of our knowledge, such a long-term ($\sim$ 46 yr) spin evolution study of a wind-fed X-ray pulsar is being reported for the first time. It should be noted that the spin period evolution over periods of $\sim$10 to 20 yr has been studied earlier in Cen X-3 \citep{tsunemi1989pulse} and LMC X-4 \citep{molkov2016near} which are both disc-fed pulsars. We use all the spin period measurements reported in literature and infer new spin periods using archival data from the \textit{Rossi X-ray Timing Explorer} mission (\textit{RXTE}, \citet{bradt1993rxte}  and the currently operational \textit{AstroSat} mission \citep{agrawal2006broad}. 

\subsection{\textit{RXTE} observations and data reduction}
The \textit{Rossi X-ray Timing Explorer} (\textit{RXTE}) was operational for about 16 yr from 1996 February until 2012 January. During its lifetime it observed Vela X-1 numerous times and about a hundred archival \textit{RXTE} observations are available for
this star, with each observation ranging from about a thousand seconds to $\sim$ 40 ks.
We have analysed some of these archival observations to obtain the spin period of Vela X-1 at different epochs. The log of \textit{RXTE} observations used in our study are shown in Table ${\ref{t1}}$. Three science instruments were flown onboard this mission viz. 
the Proportional Counter Array (PCA, \citet{jahoda1996euv}), the High Energy Timing Experiment (HEXTE; \citet{rothschild1998flight})
and the All Sky Monitor (ASM; \citet{levine1996first}). In our present
work, we have used archival data from the PCA instrument. The PCA consists of five co-aligned Xenon proportional counter units (PCUs) with total effective area of about $6500 ~\rm{cm^2}$ and sensitive in the energy range
from $2 ~\rm{keV}$ to $\sim 60 ~\rm{keV}$ \citep{jahoda1996euv}. We used data only from the PCU2 as it was functional most of the time during the \textit{RXTE} lifespan, while other PCUs suffered occasional breakdowns. 
We analysed the \textit{RXTE} data using the standard NASA \texttt{HEASOFT} software package (version 6.12) released by the High Energy Astrophysics Science Archive
Research (HEASARC) Center.\footnote{\url{http://heasarc.gsfc.nasa.gov/}} 
The detailed procedure of data analysis using \texttt{FTOOLS} are illustrated
in \textquotedblleft The \textit{RXTE} Cook Book: Recipes for Data Analysis and Reduction \textquotedblright
\footnote{\url{https://heasarc.gsfc.nasa.gov/docs/xte/recipes/cook\_book.html}} and \textquotedblleft {The ABC of XTE} \textquotedblright \footnote[3]{\url{http://heasarc.gsfc.nasa.gov/docs/xte/abc/contents.html}}, written and maintained
by \textit{RXTE} Guest Observer Facility (\textit{RXTE} GOF).\\

The X-ray light curves with $125~ \rm{ms}$ resolution were extracted using standard \textit{RXTE} analysis software \texttt{FTOOLS} ver. 6.13.
All the light curves were rebinned in 1.0 s bins before further processing. When determining the source ``goodtime'' intervals, the following data selection criteria were used
for the extraction of source counts. To avoid possible contamination due to X-rays from the Earth's limb, data was extracted
only when the satellite was pointing more than $10^{\circ}$ above the horizon. To avoid possible contamination from activation in the detectors due
to the high particle rates in the SAA (South Atlantic Anomaly) passages, data were rejected from a 40 min interval beginning with the satellite entering the SAA. It was also
ensured that the satellite pointed within $0.02^{\circ}$ of the source position. 
Some of the observations had low exposure time ($\rm{\sim 3000 ~s}$) and so they were merged with other close observations to generate a light curve with long 
exposure. In order to remove the effect caused by the motion of the satellite and the Earth, the observed times of arrival of the pulse were converted to those of the barycentre of the solar system. The barycentric corrections to the event data files were applied using the \texttt{FTOOLS} utility \lq \texttt{FAXBARY}\rq. ~The light curves were rebinned to 16 s to enhance the signal-to-noise ratio (SNR) before periodicity searches.
The resulting light curve has data gaps in between and so we used the method of Scargle \citep{scargle1982} for periodicity searches in \textit{RXTE} data.  The derived pulse periods are tabulated in the appendix $\ref{appendix:a}$.

\subsection{LAXPC observations and data reduction}
We have analysed observations from the Large Area X-ray Proportional Counter (LAXPC) instrument onboard the \textit{AstroSat} mission \citep{agrawal2006broad}. The \textit{AstroSat}/LAXPC observations were carried out on 2015 November 25 and 2015 November 26 covering orbits 869-874, during the Performance Verification (PV) phase of \textit{AstroSat}. We have also analysed the Target of Opportunity (ToO) observations of Vela X-1 from 2019 March 12 to 2019 March 16 covering 13 orbits available on the ISRO Science Data Archive for the \textit{AstroSat} mission \footnote[4]{\url{https://www.issdc.gov.in/astro.html}}. The log of \textit{AstroSat} observations used in our study is shown in Table ${\ref{t2}}$.\\

\begin{figure}
\centering
  \includegraphics[width=\columnwidth]{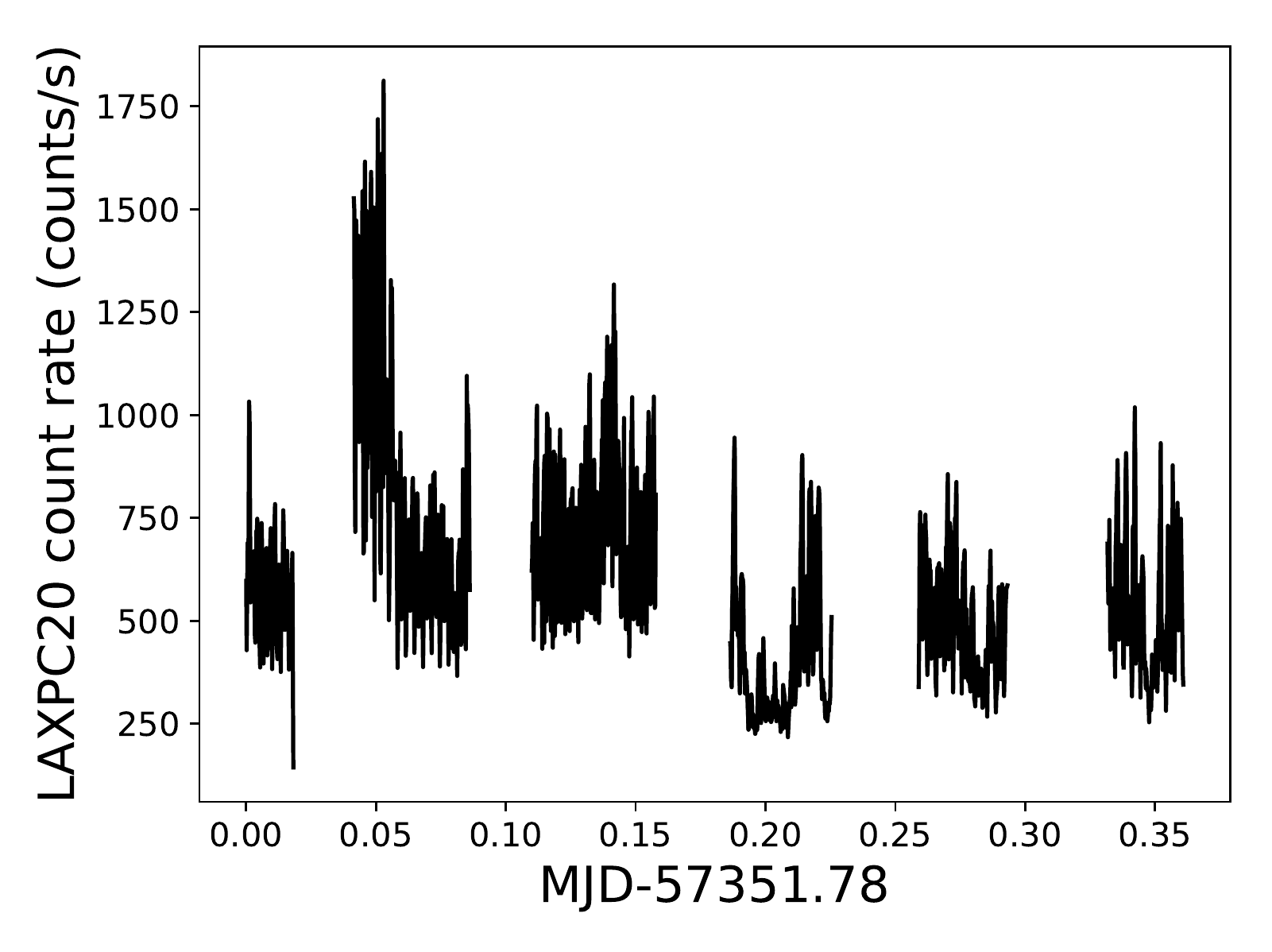} 
  \caption{Light curve of Vela X-1 in the 3-80 keV energy band using \textit{AstroSat} data from orbit 869-874. The light curve has been rebinned to 16 s. The gaps in the light curve are due to the passage of the satellite through the South Atlantic Anomaly regions.}
 \label{f1}
\end{figure}

LAXPC consists of 3 identical collimated detectors (LAXPC10, LAXPC20 and LAXPC30), having 5 anode
layer geometry with 15 cm deep X-ray detection volume providing an effective area of about 4500 cm$^2$ at 5 keV, 6000 cm$^2$ at 10 keV and about 5600 cm$^2$ at about 40 keV  \citep{roy2019laxpc,chandra2020study}.
The arrival times of X-ray photons are recorded with a time resolution of 10 $\rm{\mu}$s.
The details of the characteristics of the LAXPC instrument are available in \citep{yadav2016large,agrawal2017large,roy2016}. The calibration details of LAXPC instrument are given in \citet{antia2017calibration}. We have used softwares available from the \textit{AstroSat} Science Support Cell \footnote[5]{\url{http://astrosat-ssc.iucaa.in/?q=laxpcData}} to reduce Level-1 data to Level-2 data. Level-2 data contains (i) light curve in broad band counting mode (modeBB) and (ii) event mode data (modeEA) with information about arrival time, pulse height and layer of origin of each detected X-ray and (iii) housekeeping data and parameter files are stored in mkf file. We have used laxpc software tool having single routine to extract spectra, light curve and background, \textquotedblleft \texttt{LAXPCSOFT} \textquotedblright ~to extract light curves having 1 s resolution using the event mode data. Fig. $\ref{f1}$ shows light curve of Vela X-1 in the 3-80 keV energy band obtained using LAXPC20 data from orbit 869-874. 

We correct X-ray photons arrival times to the solar system barycentre using the \textit{AstroSat} barycentric correction utility \lq \textit{as1bary}\rq. The orbit files for barycentric correction are generated using \textit{AstroSat} orbit file generator
\footnote[6]{\url{http://astrosat-ssc.iucaa.in:8080/orbitgen/}}. \lq \textit{as1bary}\rq ~requires \texttt{HEASOFT} software package (version 6.17 or higher) and so we have used the latest \texttt{HEASOFT} software package (version 6.26) for our analysis. The light curves were rebinned to 16 s to enhance the signal-to-noise ratio (SNR) before searches for periodicity. 
We use the \texttt{FTOOLS} subroutine \textit{efsearch} to obtain the best estimated pulse periods. The inferred pulse periods are shown in the appendix $\ref{appendix:a}$.\\

\begin{table*}
\caption{Log of \textit{RXTE} PCU2 observations used in this study.} 
\label{t1}
\centering 
\begin{tabular}{c c c c c c} 
\hline\hline 
S. no. & ObsID & Time of Observation & Exposure (s) & MJD (mid) & Useful exposure (s) \\ [0.5ex] 
\hline 
1 & 10141-01-02-00 & 1996-02-22 02:15:14.7 & 6567 & 50135.16 & 5952\\ 

2 & 10144-02-01-00 & 1996-03-11 15:36:59.6 & 9956 & 50153.74 & 9232 \\ 
3 & 10142-01-01-00 & 1996-07-27 16:31:40.2 & 31953 & 50291.86 & 16176 \\ 
4 & 30102-01-01-00 & 1998-01-21 19:20:41 & 40833 & 50834.98 &
15744 \\ 
5 & 40073-01-01-03 & 2000-02-03 05:04:25.8 & 2167 & 51577.30 & 5936 \\ 
  & 40073-01-01-02 & 2000-02-03 06:34:33.6 & 2212 &  &  \\ 
  & 40073-01-01-01 & 2000-02-03 08:13:07.7 & 2203 &  & \\ 
6 & 40073-01-01-04 & 2000-02-03 22:37:04.2 & 3951 & 51578.33 & 27632 \\ 
  & 40073-01-02-00 & 2000-02-04 00:00:06.9 & 10677 &  &  \\ 
  & 40073-01-02-01 & 2000-02-04 05:10:02.8 & 1373 &  &  \\ 
  & 40073-01-02-02 & 2000-02-04 06:36:04.3 & 1971 &  &  \\ 
  & 40073-01-02-03 & 2000-02-04 08:12:04.6 & 2090 &  &  \\ 
  & 40073-01-02-04 & 2000-02-04 09:48:04.9 & 16490 &  &  \\ 

7 & 40073-01-03-00 & 2001-05-14 20:03:16.7 & 14154 & 52044.12 & 23616 \\ 
  & 40073-01-03-01 & 2001-05-15 03:47:05.3 & 12225 &   &  \\ 
  
8 & 90083-01-01-05 & 2005-01-01 00:08:48.8 & 1938 & 53371.22 & 18880 \\ 
 & 90083-01-01-06 & 2005-01-01 01:20:03.8 & 2149 &  &  \\ 
 & 90083-01-01-07 & 2005-01-01 02:51:00.9 & 13735 &  &  \\ 
 & 90083-01-01-10 & 2005-01-01 09:08:58.3 & 3779 &  &  \\ 
 9 & 90083-01-01-08 & 2005-01-02 02:52:52.3 & 13330 & 53372.50 & 32560 \\ 
  & 90083-01-01-12 & 2005-01-02 08:44:59.7 & 4000 &  &  \\ 
  & 90083-01-01-04 & 2005-01-02 10:20:00.4 & 17275 &  &  \\ 
  & 90083-01-01-13 & 2005-01-02 18:27:57.9 & 2217 &  &  \\ 
  & 90083-01-01-14 & 2005-01-02 20:05:00.4 & 2395 &  &  \\ 

10 & 90083-01-01-15 & 2005-01-03 00:54:19 & 2237 & 53373.52 & 33712 \\ 
 &  90083-01-01-16 &  2005-01-03 02:30:29.7 &  2396 &  &  \\ 
 &  90083-01-01-25 &  2005-01-03 03:47:29.5 &  3477 &  &  \\ 
 &  90083-01-01-27 &  2005-01-03 05:22:29.3 &  3417 &  &  \\ 
 &  90083-01-01-28 &  2005-01-03 06:51:29.7 &  3777 &  &  \\ 
 &  90083-01-01-17 &  2005-01-03 08:26:29.5 &  2997 &  &  \\ 
 &  90083-01-01-02 &  2005-01-03 09:56:00.1 &  17155 &  &  \\ 
 &  90083-01-01-29 &  2005-01-03 18:03:00.6 &  2573 &  &  \\ 
 &  90083-01-01-18 &  2005-01-03 19:41:01 &  2514 &  &  \\ 
 &  90083-01-01-19 &  2005-01-03 22:55:38.8 &  1777 &  &  \\
 
 11 & 90083-01-01-20 & 2005-01-04 01:59:05.3 & 2504 & 53374.54 &  30192\\ 
  &  90083-01-01-21 &  2005-01-04 03:18:00.9 &  7195 &  &  \\ 
 &  90083-01-01-30 &  2005-01-04 06:24:01.2 &  16000 &  &  \\ 
 &  90083-01-01-03 &  2005-01-04 11:17:01 & 20034 &   &  \\ 
 &  90083-01-01-22 &  2005-01-04 22:30:52.7 &  3000 &  &  \\ 

12 &  90083-01-01-23 &  2005-01-05 01:29:52.2 &  14531 & 53375.54 &  26160\\ 
 &  90083-01-01-24 &  2005-01-05 07:35:03.3 &  4000 &  &  \\ 
 &  90083-01-01-01 &  2005-01-05 09:11:01 &  14397 &  &  \\ 
 13 &  90083-01-02-01 &  2005-01-09 05:18:41.2 &  8195 &  53379.29 & 6896 \\ 
 
 14  &  93039-02-01-08 &  2007-12-03 02:56:10.2 &  995 &  54437.39 &  33600\\ 
 &  93039-02-01-03 &  2007-12-03 04:12:18.1 &  2179 &  &  \\ 
 &  93039-02-01-10 &  2007-12-03 05:15:33.7 &  16009 &  &  \\ 
 &  93039-02-01-09 &  2007-12-03 11:34:03.1 &  3833 &  &  \\ 
 &  93039-02-01-04 &  2007-12-03 13:16:18.3 &  17662 &  &  \\

15 & 93039-01-01-00 &  2007-12-07 11:24:19.9 &  17361 &  54442.42 & 67264 \\ 
 &  93039-01-01-11 &  2007-12-07 19:53:23.1 &  8289 &  &  \\ 
  &  93039-01-01-06 &  2007-12-08 00:15:02.9 &  2989 &  &  \\ 
 &  93039-01-01-09 &  2007-12-08 01:49:19.5 &  3317 &  &  \\ 
 &  93039-01-01-04 &  2007-12-08 03:09:17.3 &  20425 &  &  \\ 
 &  93039-01-01-05 &  2007-12-08 11:31:03.4 &  952 &  &  \\ 
 &  93039-01-01-01 &  2007-12-08 12:43:19.8 &  23777 &  &  \\ 
 &  93039-01-01-07 &  2007-12-08 23:46:02.8 &  3049 &  &  \\ 
 &  93039-01-01-10 &  2007-12-09 01:21:19.9 &  3437 &  &  \\ 
 &  93039-01-01-02 &  2007-12-09 02:44:16.5 &  16883 &  &  \\

16 & 93039-01-02-02 & 2008-11-25 04:22:05.7 & 18403 & 54795.52 & 29184 \\ 
 & 93039-01-02-08 & 2008-11-25 11:36:03.2 & 2451 &  &  \\ 
 & 93039-01-02-05 & 2008-11-25 13:15:13.5 & 5551 &  &  \\ 
 & 93039-01-02-03 & 2008-11-25 17:44:05.3 & 7430 &  &  \\ 
\hline 
\end{tabular}
\label{table:nonlin} 
\end{table*}
 
\begin{table*}
   \renewcommand\thetable{}
\contcaption{Log of \textit{RXTE} PCU2 observations used in this study.} 
\label{tab:continued}
\centering 
\begin{tabular}{c c c c c c} 
\hline\hline 
S. no. & ObsID & Time of Observation & Exposure (s) & MJD (mid) & Useful exposure (s) \\ [0.5ex] 
\hline 
 
17 & 93039-02-02-00 & 2008-11-30 01:27:03.7 & 22072 & 54801.04 & 42480 \\ 
 & 93039-02-02-01 & 2008-11-30 10:55:37.9 & 2263 &  &  \\ 
 & 93039-02-02-02 & 2008-11-30 14:01:03.6 & 8676 &  &  \\ 
 & 93039-02-02-03 & 2008-11-30 20:18:58.5 & 15540 &  &  \\ 
 & 93039-02-02-04 & 2008-12-01 02:36:03.2 & 10310 &  &  \\ 
 & 93039-02-02-05 & 2008-12-01 07:18:58.2 & 5823 &  &  \\ 
 \hline 
\end{tabular}
\label{table:nonlin} 
\end{table*}

\begin{table*}
\caption{Log of \textit{AstroSat} LAXPC observations used in this study.} 
\label{t2}
\centering 
\begin{tabular}{c c c c c} 
\hline\hline 
S. no. & Orbit & Time of Observation & MJD (start) & Useful exposure (s)  \\  
       &     & (yyyy-mm-dd hr:min:sec) &            &   \\[0.5ex]

\hline 

1 & 869 & 2015-11-25 18:37:25-2015-11-26 04:02:47 & 57351.78 & 22052  \\ 
 & 870 &  &  &   \\ 
 & 871 &  &  &   \\ 
 & 872 &  &  &   \\ 
 & 873 &  &  &   \\ 
 & 874 &  &  &   \\ 
2 & 18673 & 2019-03-12 16:31:52-2019-03-13 06:28:17 & 58554.68 & 34148  \\ 
  & 18674 &  &  &   \\ 
 & 18675 &  &  &   \\ 
 & 18676 &  &  &   \\ 
 & 18677 &  &  &   \\ 
 & 18680 &  &  &   \\ 
3 & 18716 & 2019-03-15 14:42:44-2019-03-16 03:30:05 & 58557.61 & 36697  \\ 
 & 18717 &  &  &  \\ 
 & 18718 &  &  &  \\ 
 & 18719 &  &  &  \\ 
 & 18720 &  &  &  \\ 
 & 18721 &  &  &  \\ 
 & 18723 &  &  &  \\ 

\hline 
\end{tabular}
\label{table:nonlin} 
\end{table*}

\section{Results}
\subsection{Long-term spin-down apparition in Vela X-1}

\begin{figure}
\centering
  \includegraphics[width=\columnwidth]{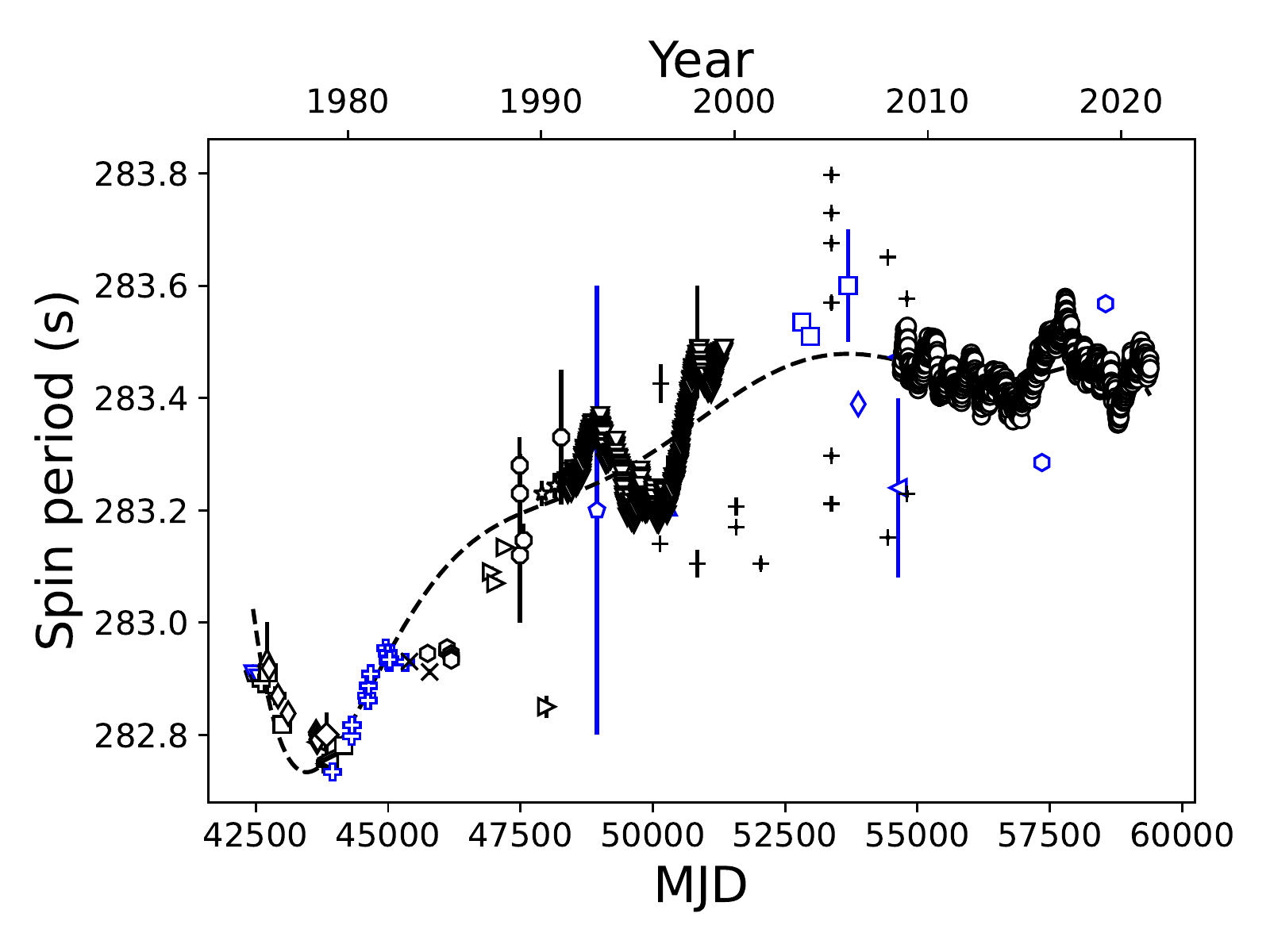} 
  \caption{Long-term spin history of Vela X-1 from 1975 February until 2021 June. The general trend of secular spin-down of the X-ray pulsar superposed on random episodes of spin-up and spin-down behaviour is discernible. The dashed curve shows the best fit polynomial to the spin evolution of Vela X-1. 
  A careful look at the long-term pulse period evolution also suggests presence of cyclic trend in the long-term spin evolution of Vela X-1. The different markers indicate spin period measurements from different observatories ($\pentagon$: \textit{Copernicus}, \textcolor{blue}{$\bigtriangledown$}: \textit{OAO 3}, \textcolor{black}{\ding{59}}: \textit{SAS 3}:, $\bigtriangleup$: \textit{Ariel V}, $\square$: \textit{COS B}, $\diamond$: \textit{OSO 8}, $\lhd$: \textit{HEAO 1}, $\Diamond$: AIT/MPI balloon X-ray detector, \textcolor{blue}{\ding{59}}: \textit{Hakucho}, x: \textit{Tenma}, $\hexagon$: \textit{EXOSAT}, $\rhd$: \textit{Ginga}, $\octagon$: \textit{KVANT}, *: \textit{GRANAT}, \textcolor{blue}{$\pentagon$}: \textit{ROSAT}, $\bigtriangledown$: \textit{CGRO}/BATSE, +: \textit{RXTE}, \textcolor{blue}{$\bigtriangleup$}: \textit{BeppoSAX}, 
  \textcolor{blue}{$\square$}: \textit{INTEGRAL}, \textcolor{blue}{$\diamond$}: \textit{XMM-Newton}, \textcolor{blue}{$\lhd$}: \textit{Suzaku}, X: \textit{NuSTAR}, $\Circle$: Fermi/GBM, \textcolor{blue}{$\hexagon$}: \textit{AstroSat}). The symbols are shown in different colours for the various missions. The dark shaded patches are due to close cadence of BATSE and \textit{Fermi} observations such that they almost overlap
showing dark patches.}
 \label{f2}
\end{figure}

Making use of all the spin periods of Vela X-1 reported in literature as well as the values derived from our analysis of \textit{RXTE}/PCA and \textit{AstroSat}/LAXPC, we construct the long-term spin history of Vela X-1 which is shown in Fig. {\ref{f2}}.

\begin{figure}
\centering
  \includegraphics[width=\columnwidth]{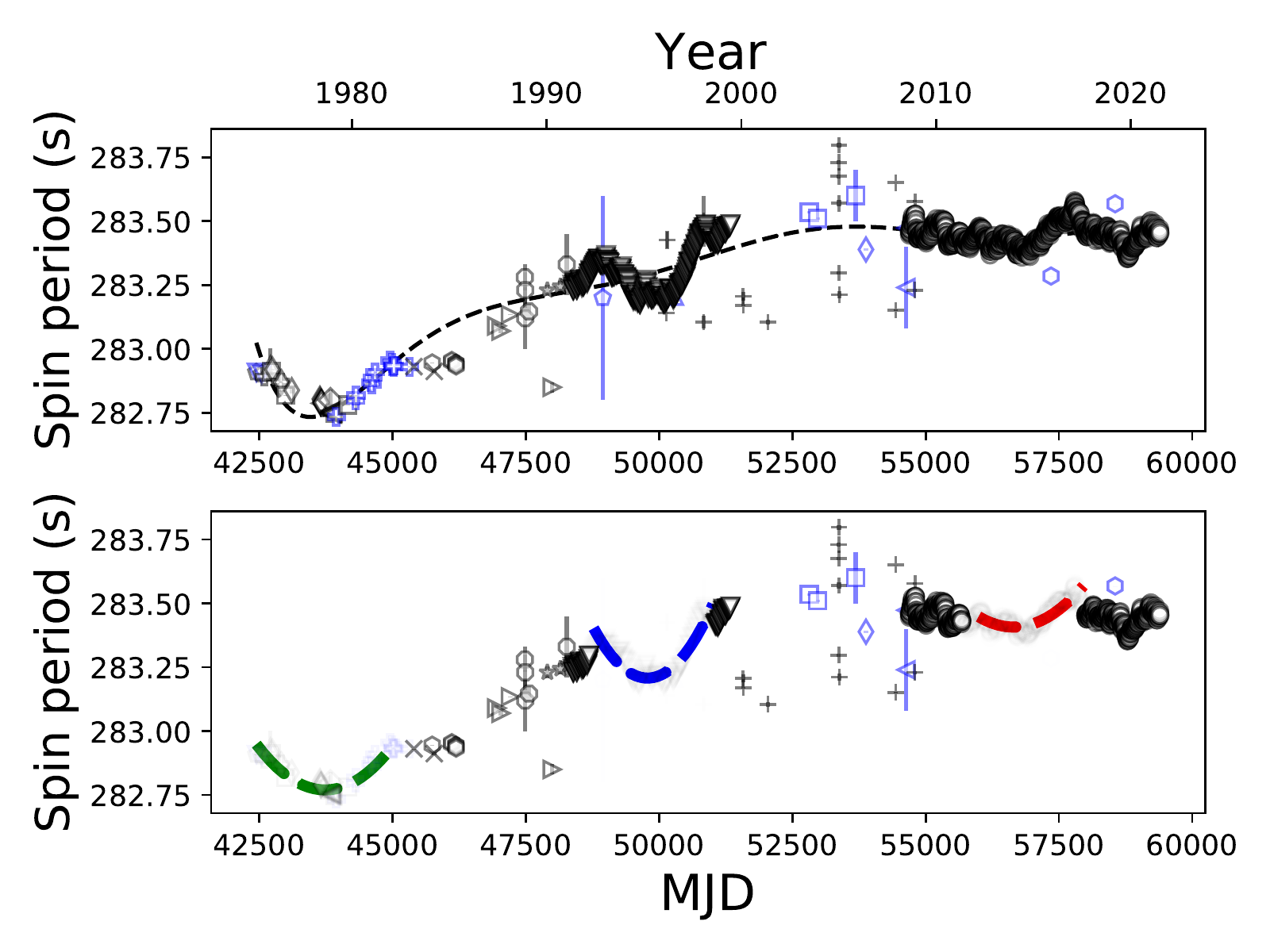} 
  \caption{Long-term spin history of Vela X-1 from February, 1975 until June, 2021 (upper panel, same as shown in Fig. \ref{f2}). Three distinct episodes of torque reversals  from spin-up to spin-down occur around MJD 44000, 50000 and 57000 which have been fitted using a parabola shown in dashed green, blue and red curve respectively (bottom panel). The radius of curvatures of the parabolas fitted to the torque reversals around these epochs are different and show asymmetricity.}
 \label{f3}
\end{figure}

The spin history spans a period of 46 yr (1975 February until 2021 June) and this is probably the longest time period over which the spin period for an accretion powered pulsar has been measured.
So far, long-term trends in spin period have been observed only in X-ray pulsars harbouring an accretion disc \citep{tsunemi1989pulse,gonzalez2012spin,molkov2016near,vasilopoulos2019ngc}. The 
remarkable long-term secular spin-down trend of Vela X-1 is clearly discernible, albeit it shows random alterations between spin-up and spin-down regimes on shorter time-scales, which is
the paradigm of wind-fed pulsars \citep{deeter1989vela,bildsten1997observations}.  Vela X-1 was found to be on a spin-up evolutionary track after its discovery in 1975. This trend continued for four years until 1979 and then Vela X-1 experienced an abrupt transition to spin-down regime, which lasted about
three years. It is intriguing to note that the culmination of this spin-down phase has spin period value comparable to that when its spin period was first measured in 1975. 
Vela X-1 continued its spin adventure and thereafter switched into a bizzare mode, wherein the pulse period was more or less stable on time-scale of about 3-4 yr. This almost constant spin period phase is really intriguing for a wind-fed pulsar as wind-fed accretion powered pulsars are known to show random variations in spin period
on short time-scales (as short as within a few days, \citet{boynton1984new,boynton1986vela,deeter1989vela,baykal1993empirical,de1993simple}). Thereafter, the pulsar switched to a remarkable long-term spin-down behaviour (with superposed sporadic episodes of spin-up and spin-down behaviour on relatively shorter time-scales) lasting almost three decades or so. A careful
examination of the Fig. \ref{f2} suggests that the rate of spin-up and spin-down can vary significantly on various time-scales during the entire recorded spin history of Vela X-1. We observe dips in spin period around MJD 44000, 50000, 52000, 54000, 57000 and 59000 which hints that there might be cyclic transitions between spin-up and spin-down on
time-scales of about a few thousand days or so. Fig. \ref{f3} shows the spin evolution of Vela X-1 for about 46 yr and we identify three distinct episodes of torque reversals from spin-down to spin-up regimes around MJD 44000, 50000 and 57000. These torque reversals occur nearly in a cyclic manner after about 6000-7000 d ($\sim$ 16.5-19.4 yr) which is really intriguing for a wind-fed pulsar which are known to show random torque reversals on much smaller time-scales. Interestingly, the radius of curvatures of the parabolas fitted to the spin evolution around these epochs are different and asymmetric. This suggests that the pulsar spends different durations in these states and the rate of spin-up and spin-down are markedly different during these torque reversals.  

\subsection{Nearly periodic spin period reversals in Vela X-1 on long time-scales }

\begin{figure}
\centering
  \includegraphics[width=\columnwidth]{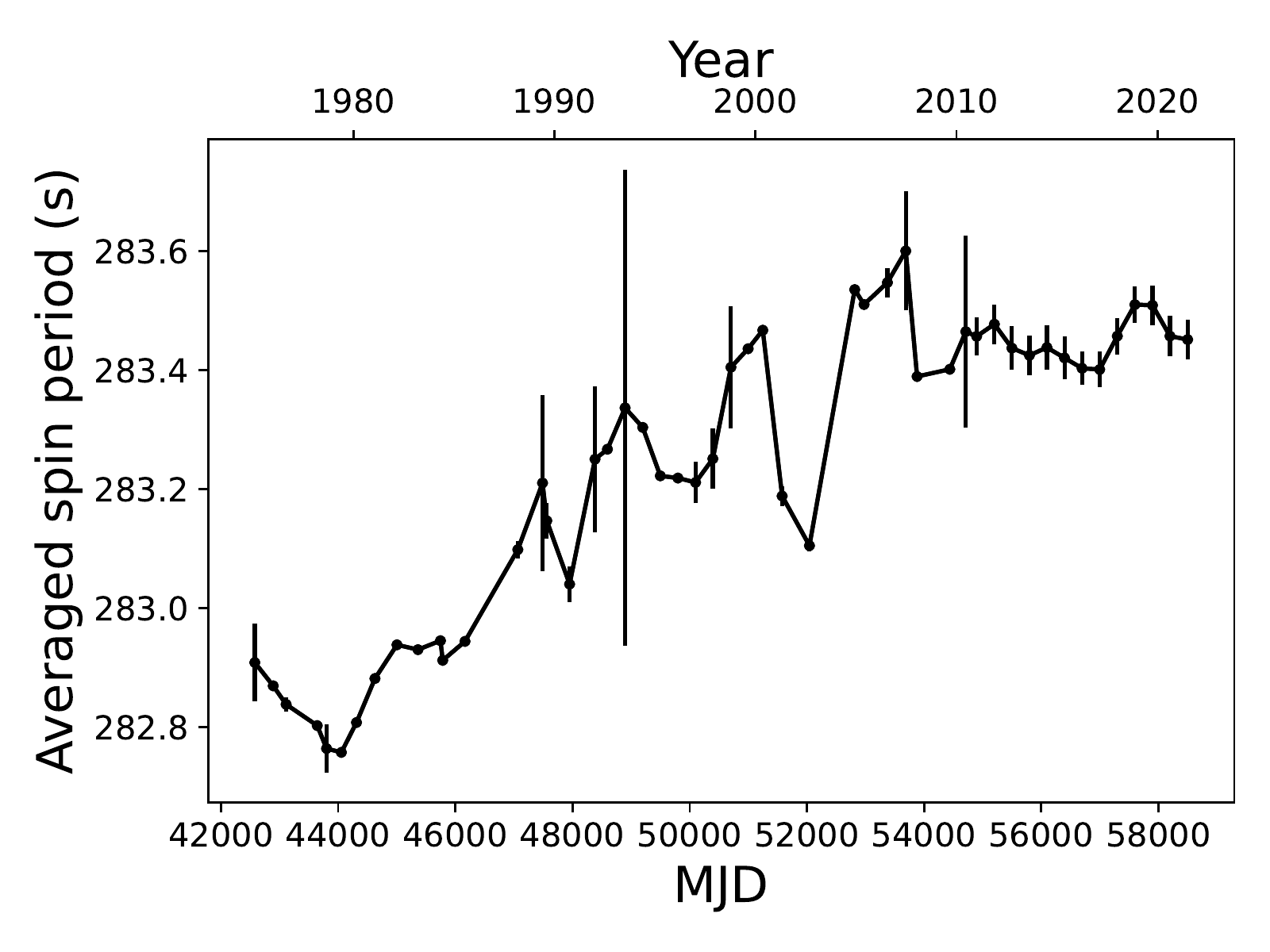} 
  \caption{Long-term averaged spin history of Vela X-1 from 1975 February until 2021 June. The spin periods have been averaged using a time window of 300 d. The presence of quasiperiodic episodes of spin-up and spin-down (on time-scales of about 2000 d) superposed on long-term spin-down trend stands out.}
 \label{f4}
\end{figure}

We further probe possible cyclic variations in the spin period of Vela X-1 by constructing the long-term averaged spin history using a smoothing window of 300 d (Fig. \ref{f4}). 

\begin{figure}
\centering
  \includegraphics[width=0.8\linewidth,angle =-90]{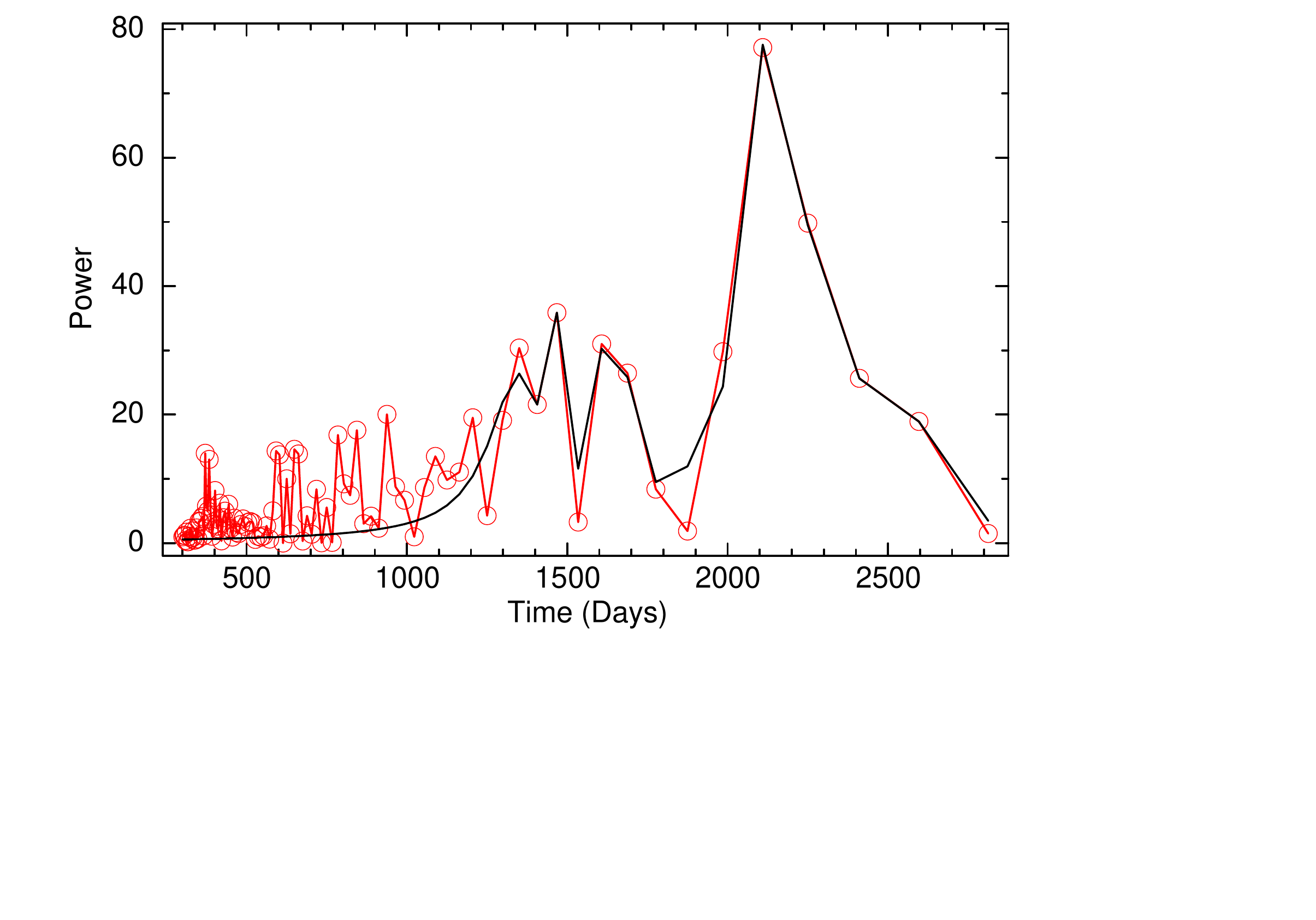} 
  \caption{Lomb-Scargle periodogram of the long-term spin evolution of Vela X-1 shown earlier in Fig. \ref{f2}. The periodogram peaks at 2154.58 d (about 5.9 yr). The detected period is found to have more than 5$\sigma$ confidence.}
 \label{f5}
\end{figure}

This averages out random transitions in spin period on short time-scales and clearly brings out the nearly cyclic variations in spin period on long time-scales. We observe prominent dips in the spin period around MJD 44000, 48000, 50000, 52000, 54000, 57000 and 59000. It is really intriguing that the long-term averaged spin variations in Vela X-1 is almost cyclic having periodicity of about 2000 d. The tiny dip in spin period around MJD 46000 is likely an artefact of smoothing. 
Besides, the rate of spin-up and spin-down varies significantly and quite differently during each of these
slow transitions in spin period as is evident from the asymmetric nature of these gradual spin period changes occurring on long time-scales. These subtle observations gleaned from averaged spin history in Vela X-1 can be helpful in probing the underlying phenomena causing nearly cyclic spin changes in this pulsar on time-scale of about 2000 d. In case the spin-down rate is smaller than the spin-up rate during 
these gradual transitions (e.g. notice transitions in spin period around MJD 48000, 50000 and 52000), it may suggest a phenomena involving some kind of relaxation effect such as a complex interplay of slow changes in the accretion rate of the neutron star, internal torques acting on the crust of the neutron star and/or magnetospheric state variations of the neutron star. 
We notice from Fig. \ref{f2} that the cyclic modulation in spin period is superposed on a long-term spin-down trend (starting around MJD 44000 until around MJD 53000). Fig. \ref{f5} shows the Lomb-Scargle periodogram of the long-term spin evolution of Vela X-1. We obtain the time-scale of nearly cyclic spin period evolution to be 2154.58 d ($5.90^{+0.05}_{-0.10}$ yr). Table $\ref{t3}$ lists salient parameters of X-ray pulsars in which nearly cyclic spin period evolution on time-scales of years has been detected.

\begin{table*}
\caption{List of known X-ray pulsars which show nearly periodic spin period changes on long time-scales.} 
\label{t3}
\centering 
\begin{tabular}{c c c c c c} 
\hline\hline 
 Name of X-ray pulsar & Spin period (s) & Orbital period (d) & Spin change periodicity (yr) & wind/disc-fed & Reference\\  

\hline 
 Cen X-3 &  $\sim 4.8$ & $\sim 2.1$ & $\sim 9.2$ & disk-fed & 1\\
 LMC X-4 &  $\sim 13.5$ & $\sim 1.4$ & $\sim 6.8$ &  disk-fed & 2\\  
 Vela X-1 &  $\sim 283$ & $\sim 8.9$ & $\sim 5.9$ & wind-fed & 3 (this work)\\
\hline 
\end{tabular}
\label{table:nonlin} 
\\(1) \cite{tsunemi1989pulse}, (2) \cite{molkov2016near}
and (3) this work.
\end{table*}

\section{Discussions}

From the spin period measurements over a period of 46 yr, one can discern an overall trend of spin-down of the source though it will be noticed from Fig. \ref{f2} that since MJD 51000 the spin period has largely stayed at $\sim$283.4 s with erratic episodes of spin-up and spin-down. The results in the previous section show nearly periodic spin
period variations in Vela X-1 on time-scales of about 5.9 yr which is superposed on long-term spin-down trend since its discovery in 1975. 
We also find nearly cyclic turnover from spin-up to spin-down regimes on decadal time-scales of about 17-19 yr (three cycles clearly identified in about 46 yr of spin period evolution of this X-ray pulsar) which, to the best of our knowledge, has been detected in any accretion powered pulsar for the very first time. In the following section, we discuss possible mechanisms which can lead to nearly cyclic spin period variations on time-scales of years in wind-fed pulsars like Vela X-1. We would like to emphasize that as the temporal variation of accretion torques in wind-fed accretion powered pulsars is complex and depends on many parameters such as the stellar wind velocity, stellar wind density, magnetospheric state of the pulsar, energetics of the plasma flow in the vicinity of the neutron star etc. a combination of the following mechanisms may operate simultaneously in this wind-fed X-ray pulsar.

\subsection{Cyclic mass loss in the donor star HD 77581?}

Variations in the spin period of an accretion powered pulsar is the manifestation of the interplay of two torques acting on the compact object viz. the accretion torque exerted 
by the accreting matter captured from the stellar wind of the donor star (for a wind-fed pulsar such as Vela X-1) and the internal torque due to the dynamic coupling between the solid crust and the superfluid matter in the core of the neutron star \citep{nagase1989accretion}. To decouple these two torques acting on a neutron star is
a challenging exercise. It is believed that spin wandering on short time-scales can be caused either by variations in the accretion torque \citep{elsner1976accretion,ghosh1979accretion} and/or by variations in the internal torque applied by the superfluid core  \citep{lamb1978period,ghosh1979accretion} on the crust of the neutron star. It has also been suggested in some earlier studies that variations in spin period over sufficiently long time-scales is the manifestation of the external torque acting on the neutron star \citep{pringle1972accretion,lamb1973model,ghosh1979accretion}. \citet{nagase1984secular} suggest that the long-term (about 3 yr or longer) spin-down trend in Vela X-1 during 1979-81 is not caused due to the internal torque because it is less likely that the coupling between the crust and the superfluid core of the neutron star can sustain for such a long time. In addition, they also discuss about ineffectiveness of the internal torque in causing short-term, abrupt spin period changes. In the light of the above arguments, we surmise that the (external) accretion torque is the dominant torque acting on the neutron star over long time-scales (a few tens of years) and variations in this torque manifest as dynamic changes in the accretion torque and ultimately spin changes in Vela X-1 on long time-scales. Variations in this torque occur due to fluctuations in the accretion rate of matter from the stellar wind of the donor star. It is known that the stellar wind from early-type stars vary irregularly on time-scales from minutes to years and so it has been suggested that variations in the stellar wind from the donor star can cause reversals in the accretion torque acting on the neutron star on time-scales of days to years \citep{nagase1989accretion}. Interestingly, \citet{tsunemi1989pulse} surmise that the 9.2 yr cyclic spin period variations detected in Cen X-3 maybe caused by changes in the activity of the massive O6-8 III companion star. Variations in the stellar activity of the companion star may lead to changes in the stellar wind velocity and density. This will result in varying mass loss rate
which may modulate the accretion rate and hence correlate with spin period changes detected in Cen X-3 by \citet{tsunemi1989pulse}. Vela X-1 is a wind-fed pulsar accreting from an early-type star and therefore variations in mass-loss from the donor star will manifest as spin changes in the compact object on long time-scales. Thus, we infer that nearly periodic spin variations in Vela X-1 on long time-scales of about 5.9 yr occur possibly due to cyclic mass-loss/stellar wind variations from the donor star HD 77581. Long-term optical monitoring of the companion star on decadal time-scales might shed light on this plausible mechanism.\\

\begin{figure}
\centering
  \includegraphics[width=\linewidth]{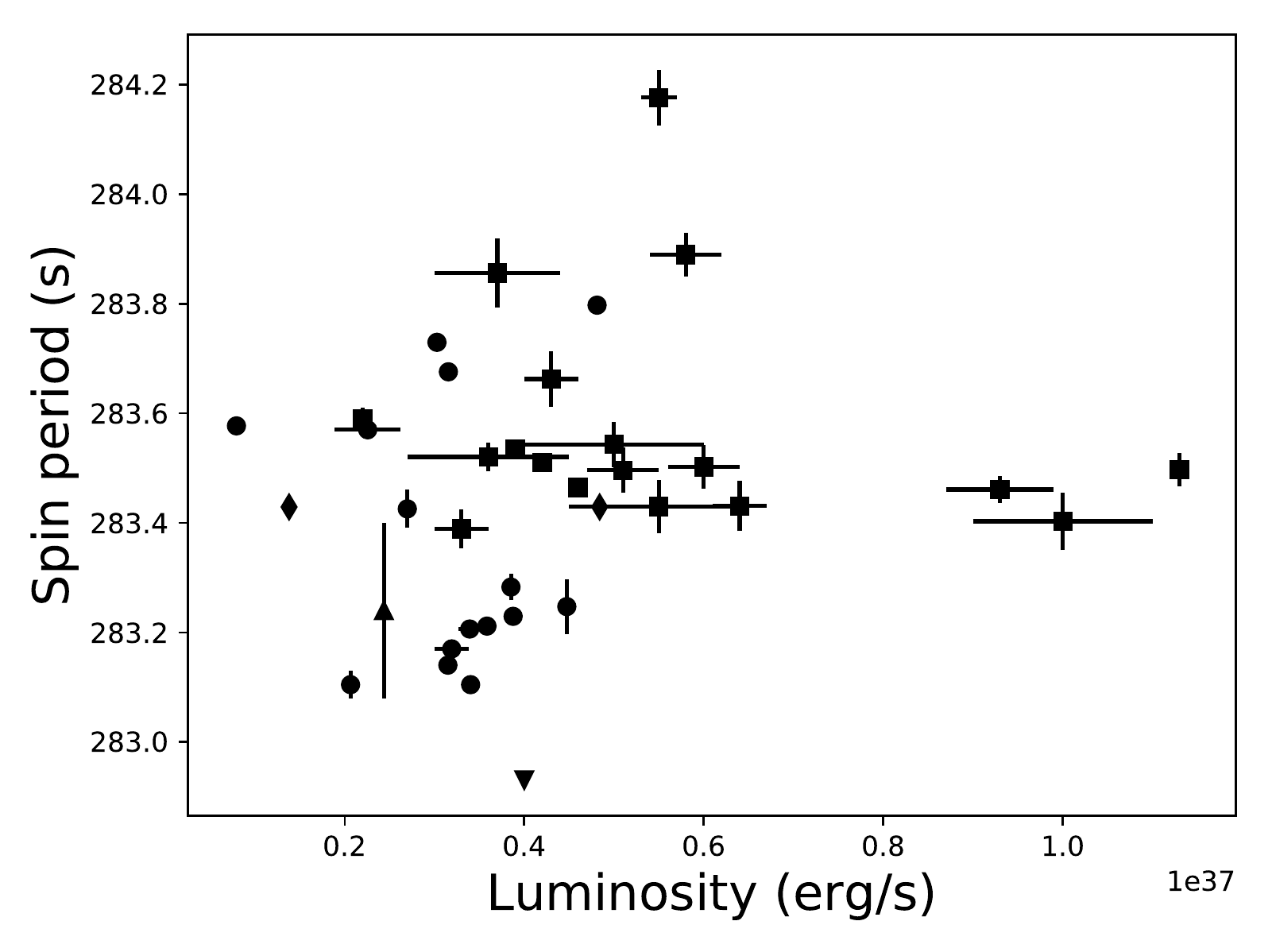} 
  \caption{Plot showing spin period vs X-ray luminosity of Vela X-1. Different symbols indicate results obtained from our work and those available in the literature. Circles indicate luminosity in the range 2-80 keV (this work), inverted triangle indicates luminosity in the range 2-30 keV \citep{nagase1986circumstellar}, squares indicate luminosity in the range 3-100 keV \citep{lutovinov2009timing},  triangle indicates luminosity in the range 0.3-70 keV \citep{maitra2013pulse} and diamonds indicate luminosity in the range 3-79 keV \citep{fuerst2013nustar}. Please note that energy ranges of luminosities are different and can be compared because the contribution to X-ray luminosity at higher energies decreases significantly.}
 \label{f6}
\end{figure}

Changes in X-ray luminosity of the compact object are a tell-tale signatures of variations in the accretion rate of captured matter from the stellar wind. We compare the luminosity and the pulse periods of Vela X-1 obtained using \textit{RXTE} PCU2 data and those reported in literature (Fig. \ref{f6}). 

It should be noted that the range of energies for comparing luminosities are not identical but still we can compare them as the 
contribution of high energy photons to the X-ray luminosity, diminishes significantly with energy. We compute Pearson correlation coefficient
for possible correlation between the spin period and X-ray luminosity and find it to be 0.2 indicating weak or no correlation between these parameters shown in Fig. \ref{f6}. It has been observed that the luminosity and pulse period in Vela X-1 are not clearly correlated \citep{malacaria2020ups}. However, some studies suggest that changes in the count rate on short time-scales can affect the spin of the neutron star on short time-scales of a few days \citep{kretschmar1997phase}. Interestingly, a correlated behaviour between count rate in 11-40 keV energy range (not affected by photoelectric absorption) and spin period has been observed approximately over 50 d or so which suggests possible correlation between accretion torque and spin period on time-scales of about a couple of months (refer Table 1 in \citet{kretschmar1997phase}). \citet{lutovinov2009timing} find using \textit{INTEGRAL} observations that the luminosity of Vela X-1 enhances almost threefold during an outburst which is accompanied by remarkable changes in the spin period and suggests that episodic enhanced accretion during flares can cause tangible changes in the spin period. Interestingly, the source spectrum and the position of cyclotron lines did not vary during this period \citep{lutovinov2009timing} which strongly suggests that the fluctuations in accretion rate was primarily responsible for sudden changes observed in the spin period during this period. However, these observed changes are detected at much shorter time-scales of about a few hours to few tens of days. Long-term X-ray monitoring of the pulsar at regular intervals using pointed mode instruments is required to investigate this premise in detail.\\

One can infer from Fig. \ref{f6} that there are variations in the pulsar period within the range $\rm{{\delta P}/P \sim 1/300}$ while the corresponding X-ray luminosity also changes by a factor of $\sim$ 3 in the range $\sim 1.5-4\times 10^{36}$\,erg\,s$^{-1}$. It has been shown by \citet{shakura2012theory,shakura2014theory, shakura2018quasi} that the quasi-spherical settling accretion regime is feasible when the X-ray luminosity $\rm{L_X} \lesssim 4\times 10^{36}$\,erg\,s$^{-1}$. At higher luminosities, the accretion proceeds in the supersonic Bondi regime. The following discussion is applicable for those states of the source when the $\rm{L_X} \lesssim 4\times 10^{36}$\,erg\,s$^{-1}$. The equilibrium period for quasi-spherical accretion is given by \citep{postnov2015spin}

\begin{equation}
 P^{*}_{eq}\sim 940[s]\mu^{12/11}_{30}\left({\frac{P_b}{10~d}}\right) \dot{M}^{-4/11}_{16}v^{4}_{8},
\end{equation}

where $\mu_{30}=\mu/10^{30} [\rm{G~cm^3}]$ is the dipole magnetic moment given by $\mu=\rm{BR^3/2}$ where R is the radius of the neutron star having typical value of 10 km, $\dot{M}_{16}=\dot{M}/10^{16}[\rm{g~s^{-1}}]$ is the accretion rate onto the neutron star, $P_b$ is the orbital period and $v_{8}=v/10^{8}[\rm{cm~s^{-1}}]$ is the characteristic stellar wind velocity. Using equation 1, we obtain

\begin{equation}
   \frac{\delta P}{P} =\frac{-4}{11} \frac{\delta L}{L} +4 \frac{\delta v}{v}. 
\end{equation}

For the Bondi-Hoyle-Lyttleton wind accretion \citep{hoyle1939effect,bondi1944mechanism,bondi1952spherically},

\begin{equation}
   \frac{\delta L}{L}=-4 \frac{\delta v}{v}+\frac{\delta{\dot{M}_o}}{\dot{M}_o},
\end{equation}

where $\dot{M}_o$ is the wind mass-loss rate from the optical companion star. Using equation 2 and 3 we obtain,

\begin{equation}
 \frac{\delta P}{P}=\frac{-4}{11} \frac{\delta{\dot{M}_o}}{\dot{M}_o}+\frac{60}{11} \frac{\delta v}{v} \sim  3\times 10^{-3}.
\end{equation}

Equation 4 suggests $\delta{\dot{M}_o}/{\dot{M}_o} \sim 15 \delta v/v$. Then from equation 2 and using ${\delta L}/L \sim 2$ (${\delta L}/L = 2(L_{max}-L_{min})/(L_{max}+L_{min})$) for the X-ray luminosity excursion, we get ${\delta v}/v \sim 2/11$ which implies from equation 3, $\delta{\dot{M}_o}/{\dot{M}_o} \sim 30/11 \sim 3$.

The mass-loss rate from the companion star has been estimated using different models and lie in the range of about $0.5 \times 10^{-6} {~\rm{M_\odot}}$ \,yr$^{-1}$ to $7 \times 10^{-6} {~\rm{M_\odot}}$ \,yr$^{-1}$ \citep{lamers1976stellar, hutchings1976stellar,conti1978stellar,dupree1980simultaneous,mccray1984spectral,sadakane1985ultraviolet,sato1986x,prinja1990photospheric,sako1999x,van2001modelling,watanabe2006x,krtivcka2012x,falanga2015ephemeris,manousakis2015stellar,gimenez2016measuring,sander2018coupling}. A detailed comparison of the mass-loss rates from the literature is given in Table 7 in \citet{kretschmar2021revisiting}. The mass-loss rate is estimated to within a factor of about $6$ comparing recent estimates from literature \citep{krtivcka2012x,falanga2015ephemeris,manousakis2015stellar,gimenez2016measuring,sander2018coupling}. The fractional mass-loss rate variations (${\delta {\dot{M}}}/ \dot{M} = 2(\dot{M}_{max}-\dot{M}_{min})/(\dot{M}_{max}+\dot{M}_{min})$) estimated from literature is about 2 using $\dot{M}_{min}=0.5 \times 10^{-6} {~\rm{M_\odot}}$ \,yr$^{-1}$ \citep{lamers1976stellar} and $\dot{M}_{max}=7 \times 10^{-6} {~\rm{M_\odot}}$ \,yr$^{-1}$ \citep{hutchings1976stellar}. Using recent estimates for $\dot{M}_{min}=0.65 \times 10^{-6} {~\rm{M_\odot}}$ \,yr$^{-1}$ \citep{sander2018coupling} and $\dot{M}_{max}=5.3 \times 10^{-6} {~\rm{M_\odot}}$ \,yr$^{-1}$ \citep{falanga2015ephemeris}, we obtain estimated fractional wind mass-loss rate variability of about 2. It should be noted that different indirect methods of the mass-loss estimate have their own systematics, which are difficult to take into account. Interestingly, the estimated fractional wind mass-loss rate variability from observations (about a factor of 2 from recent estimates) agrees to within a factor of about 1.5 from the value estimated using equation 3. This corroborates that the nearly periodic spin variations in Vela X-1 on time-scales of about 5.9 yr occur most likely due to cyclic mass-loss from the companion star HD 77581.

\subsection{Switching magnetosphere model}
We explore another mechanism which can possibly result in torque reversals in the neutron star on long time-scales. It has been known for a long time that radio pulsars (which are usually isolated systems) show unexplained stochastic deviations in their spin-down behaviour (known as ``timing noise'') on varied time-scales of a few hundred days to a few tens of years. This manifestation is akin to the random-walk behaviour in spin frequency observed in wind-fed accretion powered pulsars like Vela X-1. In an interesting study of timing irregularities of a sample of 366 pulsars, \citet{hobbs2010analysis} found some radio pulsars showing quasiperiodic structures in their long-term timing residuals. From power spectrum analysis, significant periodicities ranging from about 1.4 yr to 10 yr were found in PSR B1540-06, PSR B1642-03, PSR B1818-04, PSR B1826-17, PSR B1828-11 and PSR B2148+63 \citep{hobbs2010analysis}. Interestingly, we also find quasiperiodic variations in the long-term spin evolution of Vela X-1 on time-scales of about 5.9 yr which is comparable to those inferred for radio pulsars showing quasiperiodic changes in their timing residuals. The underlying phenomena causing quasiperiodic structures in timing noise of radio pulsars is elusive. However, it has been suggested that these changes are driven by changes in the magnetosphere of the neutron star \citep{lyne2010switched}. In this ``state-switching model'', the magnetosphere of the neutron star is suggested to harbour two or more magnetospheric states which can be stable on time-scales of years but the pulsar can switch abruptly between these states driven by changes in the parameters regulating the spin-down \citep{lyne2010switched}. This can possibly happen in Vela X-1 where the coupling between the dynamic magnetosphere and the neutron star can change in a quasiperiodic fashion. Interestingly, the disk-fed pulsar LMC X-4 has been found showing a near cyclic spin period evolution on time-scales of about 6.8 yr \citep{molkov2016near} which is within a factor of 1.2 of the inferred time-scale in wind-fed pulsar Vela X-1. Recent observations of transient pulsar V0332+53 suggests switching of coupling between the accretion disc and the neutron star magnetosphere in a disk-fed pulsar \citep{doroshenko2017luminosity}. 

\subsection{Episodic accretion from transient accretion disc in Vela X-1?}
Vela X-1 is known to accrete from the dense, strong stellar wind from its supergiant companion star (typical mass loss rate of about $\dot{M} \sim 10^{-6} {~\rm{M_\odot}}$ \,yr$^{-1}$ \citep{hutchings1974x,dupree1980simultaneous,nagase1986circumstellar,sako1999x}) which most likely does not fill its Roche lobe. However, \citet{kretschmar2021revisiting} suggest that the mass transfer in Vela X-1 might happen due to both accretion from the stellar wind all throughout the orbit and frequent episodic Roche lobe overflow near the periastron passage. Possible observational signatures of intermittent Roche lobe overflow near periastron passage might be formation of short-lived transient accretion disc around the compact object which forms and dissipates on time-scales much lesser than the orbital period of the system and manifests as sudden spin-up/down of the neutron star showing correlation or anti-correlation between spin changes and X-ray luminosity. It is believed that transient accretion disks might fuel the sparse spin-up/spin-down episodes detected by the \textit{Fermi}/GBM at irregular intervals \citep{malacaria2020ups}. The spin-up rates observed during such events are an order of magnitude higher than the spin-down rates \citep{malacaria2020ups}. However, conclusive signatures of formation of transient accretion disc in this system has evaded so far. The long-term impact of possible recurrent formation of transient accretion discs on the spin history of this pulsar is unclear. Earlier investigations of the spin evolution of the pulsar on shorter time-scales of a few days found random switches between spin-up and spin-down regimes, which is the well-known hallmark of wind-fed
pulsars \citep{boynton1984new,boynton1986vela,deeter1989vela,baykal1993empirical,de1993simple, bildsten1997observations}. However, recent numerical studies suggest formation of temporary accretion discs in wind-fed X-ray pulsars \citep{el2019formation,el2019wind,karino2019stellar} but compelling evidence of their existence has been elusive. In a recent study, \citet{liao2020spectral} infer presence of temporary accretion disc in Vela X-1 during an extended low state lasting at least 30 ks which was accompanied by an unusual spin-up event and similar Fe K $\alpha$ fluxes compared to the preceding flaring period. 

\subsection{Quasi-spherical accretion in Vela X-1}

The spin-up/spin-down variations in this pulsar can be explained using the theory of quasi-spherical accretion from the stellar wind of the companion star. The theory of quasi-spherical settling accretion has been discussed in detail by \citet{shakura2012theory} and \citet{shakura2014theory}.  According to this theory, two very diﬀerent regimes of mass accretion onto the neutron star are possible depending on the X-ray luminosity of the pulsar. For high enough luminosity of about $4\times 10^{36}$\,erg\,s$^{-1}$, the
plasma in the bow shock cools rapidly by Compton process and so the matter falls freely (supersonically) toward the
magnetosphere forming a shock in the vicinity of the magnetosphere. For lower luminosities, cooling is slightly quenched and hence the matter approaches the neutron star magnetosphere subsonically and forms a hot quasi-static shell around the magnetosphere which is referred to as the settling accretion. The accretion rate in this mode is determined by the ability of plasma to penetrate the magnetosphere via instabilities. 

Vela X-1 observations satisfy the conditions for application of this model viz. slow spin period ($\rm{P_s \sim 283 ~s}$) and X-ray luminosity of about ${4}\times 10^{36}$\,erg\,s$^{-1}$. The luminosity of Vela X-1 is about ${4}\times 10^{36}$\,erg\,s$^{-1}$ \citep{kreykenbohm2002confirmation} and occasionally exceeds this threshold during flares and outbursts. The measured surface magnetic field of the pulsar is $\rm{\sim 2.7 \times 10^{12} ~G}$ \citep{kreykenbohm2002confirmation,coburn2002magnetic} which is consistent with the expected equilibrium period of quasi-spherical accretion onto a neutron star in the regime of settling accretion \citep{postnov2015spin} given by equation 1.

In equation 1, $P^{*}_{eq}$ depends strongly on the stellar wind velocity (proportional to the fourth power of the wind velocity). Assuming that the Vela X-1 rotates about its equilibrium spin period and given that the surface magnetic field is measured using cyclotron resonance scattering features we can use it to estimate the stellar wind velocity of the matter captured by the neutron star.

\begin{equation}
 v_{8}\sim 0.57\mu^{-3/11}_{30}\dot{M}^{1/11}_{16}\left({\frac{P^{*}_{eq}/100~\rm{s}}{P_b/10~\rm{d}}}\right)^{1/4}.
\end{equation}

Using $\mu_{30}=1.35$, $\dot{M}_{16}=4.44$ (using $\mathrm{L_X}=0.1\dot{M}\mathrm{c^2}$ and $\mathrm{L_X}={4}\times 10^{36}$\,erg\,s$^{-1}$), $P^{*}_{eq}\rm{=283 ~s}$ and $P_{b}\rm{=8.96 ~d}$, we obtain $v_{8}\sim 0.8$ ($v\rm{=800 ~km~s^{-1}}$) which is approximately equal to that of $v\rm{=700 ~km~s^{-1}}$ inferred from observations \citep{gimenez2016measuring}.\\

In the following discussion, we estimate the spin-up/spin-down rates from the quasi-settling accretion theory \citep{shakura2012theory,shakura2014theory, shakura2018quasi} and compare it with the measured spin-up/spin-down rates. The estimated spin-up rate from the quasi-settling accretion theory is given by \citep{postnov2015spin}

\begin{equation}
 {\dot{\omega}^*}_{su}\sim 10^{-9} \frac{\mathrm{Hz}}{\mathrm{d}} \Pi_{su} {v_{8}}^{-4} \mu^{1/11}_{30}\dot{M}^{7/11}_{16}\left({\frac{P_b}{10~\mathrm{d}}}\right)^{-1}.
\end{equation}

We obtain ${\dot{\omega}^*}_{su} \sim 1.11 \times 10^{-7}$ Hz/d using $\Pi_{su} \sim 9$ \citep{shakura2012theory, shakura2014theory, postnov2015spin, shakura2018quasi}, $\dot{M}_{16}=4.44$, $\mu_{30}=1.35$, $v_{8}\sim 0.7$ and $P_{b}=8.96 ~\rm{d}$.

The estimated spin-down rate from the quasi-settling accretion theory is given by \citep{postnov2015spin}

\begin{equation}
 {\dot{\omega}^*}_{sd}\sim 10^{-8} \frac{\mathrm{Hz}}{\mathrm{d}} \Pi_{sd} \mu^{13/11}_{30}\dot{M}^{3/11}_{16}\left({\frac{P^*}{100~\mathrm{s}}}\right)^{-1}.
\end{equation}

\begin{figure}
\centering
  \includegraphics[width=\linewidth]{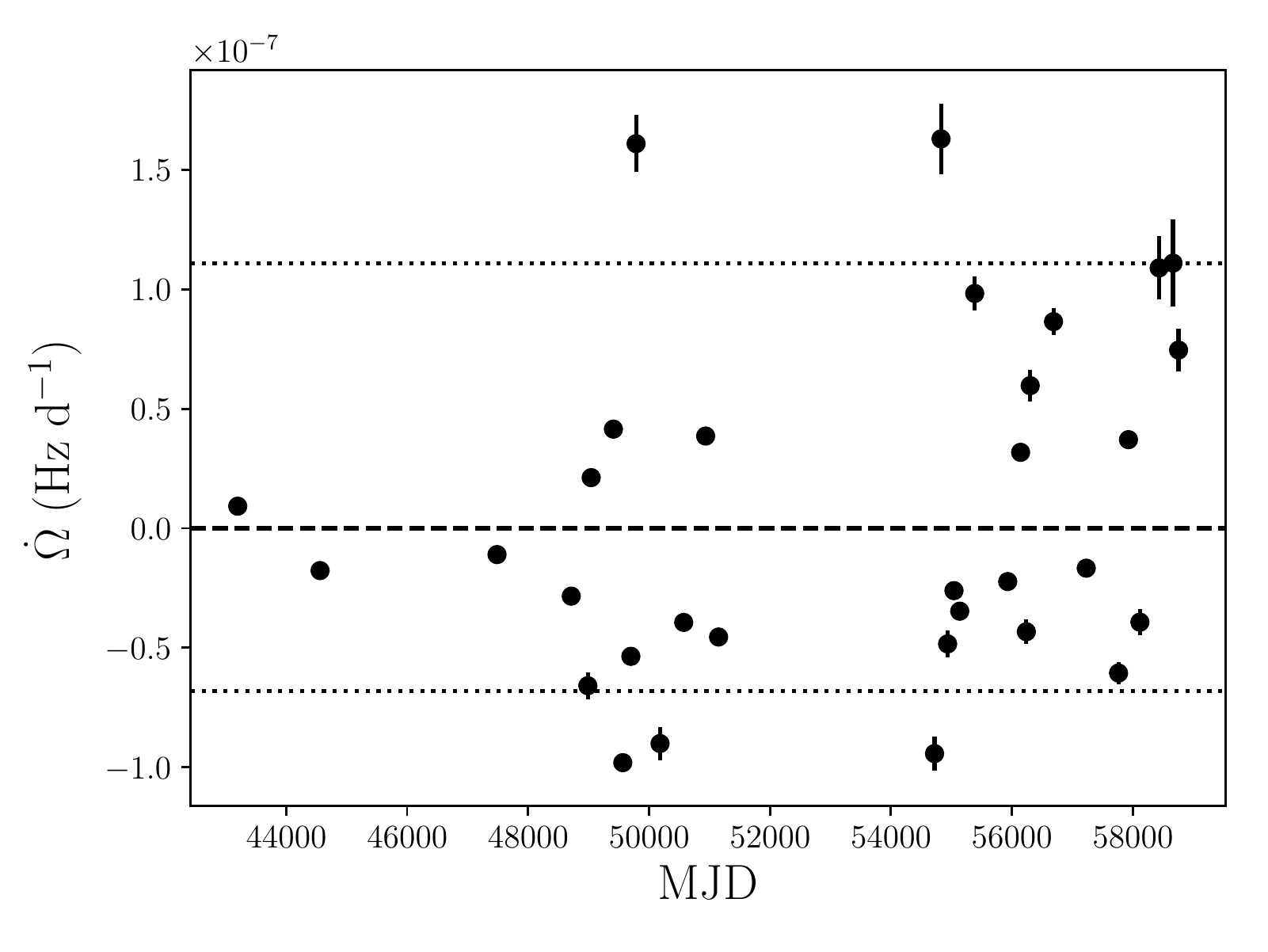} 
  \caption{Plot showing estimated spin-up and spin-down rates spread over about four decades. The horizontal dotted lines show the estimated spin-up and spin-down rates using the quasi-spherical settling accretion theory \citep{shakura2012theory,shakura2014theory}.}
 \label{f7}
\end{figure}

We obtain ${\dot{\omega}^*}_{sd} \sim -6.8 \times 10^{-8}$ Hz/d (using $\Pi_{sd} \sim 9$ \citep{shakura2012theory, shakura2014theory, postnov2015spin, shakura2018quasi}, $\dot{M}_{16}=4.44$, $\mu_{30}=1.35$, $P^{*}_{eq}\rm{=283 ~s}$ and $P_{b}\rm{=8.96 ~d}$) which is smaller than the deduced spin-up rate by a factor of about two. Fig. \ref{f7} shows the observed spin-up/down rates obtained from linear fit of \textit{BATSE} and \textit{Fermi}/GBM observations of this pulsar. It is clearly seen that the observed spin-up rate is usually higher than the spin-down rate. The average spin-up rate is $\sim 7.45 \times 10^{-8}$ Hz/d which is a factor of 1.5 lower than that estimated using the quasi-settling accretion model. The average spin-down rate is $\sim -4.64 \times 10^{-8}$ Hz/d which is also about a factor of 1.5 lower than that estimated using the quasi-settling accretion model. Thus quasi-spherical accretion model can within a factor of $\sim$2 reproduce the spin rates and may be the model that at least qualitatively explains the behaviour of Vela X-1.

\section{Summary}

We have investigated the spin period evolution of Vela X-1 over a period of about five decades and detect the long-term spin-down apparition in this pulsar. We also find random, episodic spin changes on short time-scales superposed on the long-term spin-down manifestation. To our knowledge this is the first detection of spin evolution on such a long time-scale in a wind-fed pulsar. We have also detected periodic spin period variations in Vela X-1 on time-scales of about
5.9 yr. Our study might have useful ramifications for future explorations of long-term changes in accretion history/environment in other wind-fed X-ray pulsars and help to investigate and understand the underlying phenomena causing long-term nearly cyclic spin-changes in accretion powered pulsars.

\section*{Acknowledgements}

This work is dedicated to the memory of Prof. Shashikumar Madhusudan Chitre, who sadly passed away on Monday 2021 January 11. We are extremely thankful to the reviewer for carefully going through the manuscript and making detailed, valuable and constructive suggestions which have greatly improved the presentation of this paper.
This research has made use of data obtained through the High Energy Astrophysics Science Archive Research Center (HEASARC) online service, provided by NASA/Goddard Space Flight Center. 
This publication uses the data from the \textit{AstroSat} mission of the Indian Space Research Organisation (ISRO), archived at the Indian Space Science Data Centre (ISSDC). 
We thank members of LAXPC
instrument team at TIFR and the \textit{AstroSat} project team at URSC for their contributions to the development of the LAXPC instrument. We thank the LAXPC POC at TIFR for verifying and releasing the data. \texttt{LAXPCSOFT} software is used for analysis in this paper. This research has made use of software provided by the High Energy Astrophysics Science Archive Research Center (HEASARC), which is a service of the Astrophysics Science Division at NASA/GSFC. 
This research has also made use of the \textit{Fermi}/GBM \citep{meegan2009fermi} pulsar spin evolution history provided by the \textit{Fermi} team. This research has also made use of the \textit{CGRO}/BATSE pulsar spin evolution history provided by the BATSE team. 
This research has made use of NASA's Astrophysics Data System. ADC acknowledges support of the INSPIRE scholarship of the DST, Govt. of India as an undergraduate student and computing resources in the X-ray astronomy lab, UM-DAE CEBS. ADC acknowledges support of the INSPIRE fellowship of the DST, Govt. of India. JR acknowledges ISRO for funding support and IUCAA for their facilities.

\section*{Data availability}
This research has made use of archival data from the \textit{RXTE} and the \textit{AstroSat} mission. The \textit{CGRO}/BATSE pulsar spin evolution history for Vela X-1 provided by the BATSE team is available at\\ \url{https://gammaray.nsstc.nasa.gov/batse/pulsar/data/sources/velax1.html}. The \textit{Fermi}/GBM pulsar spin evolution history for Vela X-1 provided by the \textit{Fermi} team is available at\\ \url{https://gammaray.nsstc.nasa.gov/gbm/science/pulsars/lightcurves/velax1.html}. 

\bibliographystyle{mnras}
\bibliography{mnras}

\clearpage
\onecolumn
\appendix
\section*{Appendix A: Log of all Vela X-1 pulse-period measurements used in the present work}
\label{appendix:a}
\begin{center}
\begin{longtable}{|c|c|c|c|c|c|c|}
\caption{Log of all Vela X-1 pulse-period measurements used in the present work.  Pulse period measurements from \textit{Fermi}/GBM and \textit{CGRO}/BATSE are publicly available at \url{https://gammaray.msfc.nasa.gov/gbm/science/pulsars/lightcurves/velax1.html} and \url{https://gammaray.msfc.nasa.gov/batse/pulsar/data/sources/velax1.html} respectively.}\\
\hline
{Year} & {Month} & {MJD} & {Spin period (s)} & {Uncertainty (s)} & {Satellite} & {Reference}\\ \hline
\hline
\endfirsthead
\multicolumn{7}{c}%
{\tablename\ \thetable\ -- \textit{Continued from previous page}} \\
\hline
{Year} & {Month} & {MJD} & {Spin period (s)} & {Uncertainty (s)} & {Satellite} & {Reference}\\ \hline
\hline
\endhead
\hline \multicolumn{7}{r}{\textit{Continued on next page}} \\
\endfoot
\hline
\endlastfoot
1975 &	Feb &	42448.00 & 282.90830	&       0.00340	&  \textit{Copernicus}	&  \cite{charles1978x}\\
1975 &	Feb &	42449.30 & 282.90830	&	0.00120	&  \textit{OAO 3}	&  \cite{charles1978x}\\
1975 &	May &	42551.00 & 282.90100	&	0.01100	&  \textit{Copernicus}	&  \cite{charles1978x}\\
1975 &	May &	42552.30 & 282.90100	&	0.00370	&  \textit{OAO 3}	&  \cite{charles1978x}\\
1975 &	Jul &	42600.60 & 282.89160	&	0.00040	&  \textit{SAS 3}	&  \cite{rappaport19763u}\\
1975 &	Oct &	42713.50 & 282.93700	&	0.06400	&  \textit{Ariel V}	&  \cite{charles1978x}\\
1975 &	Nov &	42728.40 & 282.91080	&	0.00120	&  \textit{COS B}	&  \cite{van1984orbital}\\
1975 &	Dec &	42750.30 & 282.91900	&	0.00300	&  \textit{OSO 8}	&  \cite{becker1978extended}\\
1976 &	May &	42900.00 & 282.87000	&	0.00400	&  \textit{SAS 3}	&  \cite{rappaport1977accretion}\\
1976 &	May &	42920.00 & 282.86900	&	0.00300	&  \textit{OSO 8}
&  \cite{becker1978extended}\\
1976 &	Aug &	42997.10 & 282.81830	&	0.00030	&  \textit{COS B}	&  \cite{van1984orbital}\\
1976 &	Nov &	43112.00 & 282.83800	&	0.01200	&  \textit{OSO 8}	&  \cite{becker1978extended}\\
1978 &	May &	43640.01 & 282.80452	&	0.00169	&  \textit{OSO 8}	&  \cite{deeter1989vela}\\
1978 &	May &	43645.62 & 282.80016	&	0.00127	&  \textit{OSO 8}	&  \cite{deeter1989vela}\\
1978 &	May &	43651.04 & 282.79188	&	0.00088	&  \textit{OSO 8}	&  \cite{deeter1989vela}\\
1978 &	May &	43654.00 & 282.78700	&	0.00400	&  \textit{HEAO 1}	&  \cite{bautz1983high}\\
1978 &	May &	43656.02 & 282.78806	&	0.00127	&  \textit{OSO 8}	&  \cite{deeter1989vela}\\
1978 &	May &	43660.75 & 282.78884	&	0.00099	&  \textit{OSO 8}	&  \cite{deeter1989vela}\\
1978 &	May &	43665.47 & 282.78899	&	0.00127	&  \textit{OSO 8}	&  \cite{deeter1989vela}\\
1978 &	May &	43669.99 & 282.79104	&	0.00110	&  \textit{OSO 8}	&  \cite{deeter1989vela}\\
1978 &	Nov &	43824.50 & 282.74860	&	0.00040	&  \textit{SAS 3}	&  \cite{rappaport1980apsidal}\\
1978 &	Nov &	43825.78 & 282.74884	&	0.00005	&  \textit{HEAO 1}	&  \cite{deeter1989vela}\\
1978 &	Nov &	43836.80 & 282.80000	&	0.04000	& 
AIT/MPI balloon X-ray detector	&  \cite{staubert1980hard}\\
1978 &	Nov &	43840.43 & 282.74606	&	0.00027	&  \textit{HEAO 1}	&  \cite{deeter1989vela}\\
1978 &	Dec &	43846.77 & 282.75233	&	0.00020	&  \textit{HEAO 1}	&  \cite{deeter1989vela}\\
1978 &	Dec &	43850.00 & 282.75130	&	0.00050	&  \textit{HEAO 1}	&  \cite{bautz1983high}\\
1978 &	Dec &	43853.10 & 282.75604	&	0.00019	&  \textit{HEAO 1}	&  \cite{deeter1989vela}\\
1978 &	Dec &	43857.59 & 282.75640	&	0.00037	&  \textit{HEAO 1}	&  \cite{deeter1989vela}\\
1978 &	Dec &	43859.94 & 282.75881	&	0.00087	&  \textit{HEAO 1}	&  \cite{deeter1989vela}\\
1978 &	Dec &	43861.38 & 282.75240	&	0.00108	&  \textit{HEAO 1}	&  \cite{deeter1989vela}\\
1978 &	Dec &	43862.75 & 282.74948	&	0.00094	&  \textit{HEAO 1}	&  \cite{deeter1989vela}\\
1978 &	Dec &	43864.21 & 282.75004	&	0.00068	&  \textit{HEAO 1}	&  \cite{deeter1989vela}\\
1978 &	Dec &	43866.59 & 282.74616	&	0.00029	&  \textit{HEAO 1}	&  \cite{deeter1989vela}\\
1978 &	Dec &	43871.00 & 282.75346	&	0.00021	&  \textit{HEAO 1}	&  \cite{deeter1989vela}\\
1979 &	Jan &	43881.64 & 282.74632	&	0.00014	&  \textit{HEAO 1}
&  \cite{deeter1989vela}\\
1979 &	Jan &	43891.30 & 282.75020	&	0.00088	&  \textit{HEAO 1}	&  \cite{deeter1989vela}\\
1979 &	Mar &	43948.10 & 282.73370	&	0.00090	&  \textit{Hakucho}      &  \cite{nagase1984secular}\\
1979 &	Oct &	44161.40 & 282.78090	&	0.00030	&   \textit{COS B}	&   \cite{van1984orbital}\\
1980 &	Mar &	44308.70 & 282.79770	&	0.00060	&  \textit{Hakucho}     & 	\cite{nagase1984secular}\\
1980 &	Mar &	44318.10 & 282.81740	&	0.00070	&  \textit{Hakucho}      & 	\cite{nagase1984secular}\\
1980 &	Dec &	44594.90 & 282.86930	&	0.00060	&  \textit{Hakucho}     & 	\cite{nagase1984secular}\\
1981 &	Jan &	44617.90 & 282.86080	&	0.00050	&  \textit{Hakucho}      & 	\cite{nagase1984secular}\\
1981 &	Jan &	44629.30 & 282.88720	&	0.00060	&  \textit{Hakucho}     & 	\cite{nagase1984secular}\\
1981 &	Mar &	44672.40 & 282.90850	&	0.00040	&  \textit{Hakucho}      & 	\cite{nagase1984secular}\\
1981 &	Dec &	44958.60 & 282.95450	&	0.00100	&  \textit{Hakucho}     & 	\cite{nagase1984secular}\\
1982 &	Jan &	44993.80 & 282.94510	&	0.00070	&  \textit{Hakucho}      & 	\cite{nagase1984secular}\\ 
1982 &	Jan &	45004.60 & 282.93150	&	0.00100	&  \textit{Hakucho}     & 	\cite{nagase1984secular}\\
1982 &	Feb &	45014.00 & 282.93500	&	0.00130	&  \textit{Hakucho}
     & 	\cite{nagase1984secular}\\
1982 &	Feb &	45023.50 & 282.92870	&	0.00220	&  \textit{Hakucho}      & 	\cite{nagase1984secular}\\
1982 &	Mar &	45031.80 & 282.93370	&	0.00120	&  \textit{Hakucho}      & 	\cite{nagase1984secular}\\
1982 &	Dec &	45320.80 & 282.92930	&	0.00050	&  \textit{Hakucho}      & 	\cite{nagase1984secular}\\
1983 &	Mar &	45404.00 & 282.93060	&	0.00030	&  \textit{Tenma}	&  \cite{nagase1984pulse}\\
1984 &	Feb &	45746.29 & 282.94500	&	0.0009	&  \textit{EXOSAT}	&  \cite{raubenheimer1990exo}\\
1984 &	Mar &	45785.00 & 282.91200	&	0.00500	&  \textit{Tenma}	&  \cite{sato1986x}\\
1985 &	Feb &	46110.06 & 282.94940	&	0.0002	&  \textit{EXOSAT}	&  \cite{raubenheimer1990exo}\\
1985 &	Feb &	46113.96 & 282.95490	&	0.0001	&  \textit{EXOSAT}	&  \cite{raubenheimer1990exo}\\
1985 &	May &	46186.07 & 282.94330	&	0.0005	&  \textit{EXOSAT}	&  \cite{raubenheimer1990exo}\\
1985 &	May &	46191.14 & 282.94410	&	0.0007	&  \textit{EXOSAT}	&  \cite{raubenheimer1990exo}\\
1985 &	May &	46193.06 & 282.93830	&	0.0003	&  \textit{EXOSAT}	&  \cite{raubenheimer1990exo}\\
1985 &	May &	46194.97 & 282.9331	&	0.0008	&  \textit{EXOSAT}	&\cite{raubenheimer1990exo}\\
1987 &	Jun &	46954.00 & 283.09000	&	0.01000	&  \textit{Ginga}	&  \cite{tsunemi1989all}\\
1987 &	Aug &	47035.00 & 283.07000	&	0.01000	&  \textit{Ginga}	&  \cite{tsunemi1989pulse}\\
1988 &	Feb &	47211.50 & 283.13400	&	0.00200	&  \textit{Ginga}	&  \cite{nagase1989accretion}\\
1988 &	Nov &	47485.00 & 283.28000	&	0.05000	&  \textit{KVANT}	&  \cite{kretschmar1997phase}\\
1988 &	Nov &	47490.00 & 283.23000	&	0.07000	&  \textit{KVANT}	&  \cite{kretschmar1997phase}\\
1988 &	Nov &	47491.00 & 283.12000	&	0.12000	&  \textit{KVANT}	&  \cite{kretschmar1997phase}\\
1989 &	Feb &	47561.00 & 283.1466	&	0.03000	&  \textit{KVANT}	&  \cite{gilfanov1989observations}\\
1990 &	Jan &	47906.00 & 283.23   	&         0.022	&  \textit{GRANAT}	&  \cite{lapshov1992two} \\
1990 &	Apr &	47995.00 & 282.85   	&         0.02	&  \textit{Ginga}	&  \cite{makishima1999cyc} \\
1990 &	Sept&	48149.00 & 283.244   	&         0.022	&  \textit{GRANAT}	&  \cite{lapshov1992two} \\
1991 &	Jan &	48269.00 & 283.33000	& 	0.12000	&  \textit{KVANT}	&  \cite{kretschmar1997phase}\\
1991 &	Feb &	48302.00 & 283.26   	&    0.013	&  \textit{GRANAT}	&  \cite{lapshov1992two} \\
1992 &	Jun &	48787.00 & 283.32600	& 	0.02000	&  \textit{GRANAT}	&  \cite{lutovinov1994timing} \\
1992 &	Nov &	48945.00 & 283.20000	& 	0.40000	&  \textit{ROSAT}	&  \cite{haberl1994rosat}\\
1996 &	Feb &	50135.16 & 283.14032	& 	0.00168	&  \textit{RXTE}	        &  present work \\
1996 &	Mar &	50153.74 & 283.42560	& 	0.03488	&  \textit{RXTE}	        &  present work \\
1996 &	Jul &	50288.00 & 283.20600	& 	0.00100	&  \textit{BeppoSAX}     &  \cite{orlandini1997bepposax}\\
1996 &	Jul &	50291.86 & 283.24722	& 	0.05048	&  \textit{RXTE}	        &  present work \\
1996 &	Aug &	50316.87 & 283.2185 	&      0.0018	&  \textit{CGRO}/BATSE   & 	\cite{bildsten1997observations} \\
1998 &	Jan &	50834.98 & 283.10471	& 	0.02481	&  \textit{RXTE}        & 	present work \\
1998 &	Jan &	50835.65 & 283.5	& 	0.1	&  \textit{RXTE}        & 	\cite{kreykenbohm2002confirmation}\\
2000 &	Feb &	51577.30 & 283.20627	& 	0.01624	&  \textit{RXTE}        & 	present work \\
2000 &	Feb &	51578.33 & 283.16998	& 	0.00378	&  \textit{RXTE}        & 	present work \\
2001 &	May &	52044.12 & 283.10471	& 	0.00980	&  \textit{RXTE}        & 	present work \\
2003 &	Jun &	52816.50 & 283.53500	& 	0.00500	&  \textit{INTEGRAL}     & 	\cite{staubert2004integral}\\
2003 &	Nov &	52976.00 & 283.51000	& 	0.00500	&  \textit{INTEGRAL}     & 	\cite{staubert2004integral}\\
2005 &	Jan &	53371.22 & 283.29715	& 	0.00589	&  \textit{RXTE}	        &  present work \\
2005 &	Jan &	53372.50 & 283.72955	& 	0.00827	&  \textit{RXTE}	        &  present work \\
2005 &	Jan &	53373.52 & 283.67569 	& 	0.00742	&  \textit{RXTE}	        &  present work \\
2005 &	Jan &	53374.54 & 283.56989	& 	0.01233	&  \textit{RXTE}	        &  present work \\
2005 &	Jan &	53375.54 & 283.79736	& 	0.00958	&  \textit{RXTE}	        &  present work \\
2005 &	Jan &	53379.29 & 283.21161	& 	0.01456	&  \textit{RXTE}	        &  present work \\
2005 &	Nov &	53688.00 & 283.60000	& 	0.10000	&  \textit{INTEGRAL}	&  \cite{schanne2007integral}\\
2006 &	May &	53881.16 & 283.389  	&         0.004	&  \textit{XMM-Newton}   & 	\cite{martinez2014accretion}\\
2007 &	Dec &	54437.39 & 283.15128	& 	0.00308	&  \textit{RXTE}	        &  present work \\
2007 &	Dec &	54442.42 & 283.65094	& 	0.00003	&  \textit{RXTE}	        &  present work \\
2008 &	Jun &	54634.00 & 283.473  	&         0.004	&  \textit{Suzaku}	&  \cite{doroshenko2011witnessing}\\
2008 &	Jun &	54634.00 & 283.24000	& 	0.16000	&  \textit{Suzaku}	&  \cite{maitra2013pulse}\\
2008 &	Nov &	54795.52 & 283.22940	& 	0.00674	&  \textit{RXTE}	        &  present work \\
2008 &	Dec &	54801.04 & 283.57703	& 	0.00452	&  \textit{RXTE}	        &  present work \\
2012 &	Jul &	56117.78 & 283.429  	&       0.0024	&  \textit{NuSTAR}	&  \cite{fuerst2013nustar}\\
2013 &	Apr &	56405.04 & 283.429  	&       0.0006	&  \textit{NuSTAR}	&  \cite{fuerst2013nustar}\\
2015 &	Nov & 57351.78	 &  283.28504  	&    0.00001	&  \textit{AstroSat}	&  present work\\
2019 &	Mar & 58554.68	 &  283.56792  	&    0.00001	&  \textit{AstroSat}	&  present work\\
2019 &	Mar & 58557.61	 &  283.61553  	&    0.00001	&  \textit{AstroSat}	&  present work\\
\hline 

\end{longtable}
\clearpage
\twocolumn
\end{center}
\bsp	
\label{lastpage}
\end{document}